\documentclass[epj]{svjour}
\usepackage[british,UKenglish,USenglish,english,american]{babel}
\usepackage{latexsym}
\usepackage{graphics}

\usepackage{etex}
\usepackage{lineno}
\usepackage[normalem]{ulem}
\usepackage{soul}
\usepackage[bookmarks,bookmarksopen,bookmarksdepth=2]{hyperref}
\hypersetup{colorlinks=true,citecolor=blue,linkcolor=blue}
\modulolinenumbers[5]

\usepackage{tikz}
\usepackage{pgfplots}
\usetikzlibrary{shapes.misc}
\usetikzlibrary{patterns,decorations.pathreplacing}

\tikzset{cross/.style={cross out, draw=black, minimum size=2*(#1-\pgflinewidth), inner sep=0pt, outer sep=0pt},
cross/.default={1pt}}
\usepackage{gensymb}
\usepackage{indentfirst}
\usepackage{fancyhdr}
\usepackage{dsfont}
\usepackage{amsfonts,amssymb,amsmath,mathtools,ulem,cancel}
\usepackage{graphicx,color,geometry,epstopdf}
\usepackage{tikz}
\usepackage{pifont}
\usepackage{epstopdf}
\usepackage{caption}
\usepackage{multirow}
\usepackage[export]{adjustbox}
\usepackage{subfig}
\usepackage{subfloat}

\graphicspath{{figures/}}

\definecolor{commenti}{rgb}{0.13,0.55,0.13}
\definecolor{stringhe}{rgb}{0.63,0.125,0.94}
\usepackage[font=footnotesize]{caption}

\usepackage{listings}
\usepackage{keyval}

\usepackage{tikz}
\usetikzlibrary{intersections}
\usetikzlibrary{calc,patterns,decorations.pathmorphing,decorations.markings}
\usepackage{pgfplots}
\usetikzlibrary{shapes.misc}
\usetikzlibrary{patterns,decorations.pathreplacing}

\tikzset{cross/.style={cross out, draw=black, minimum size=2*(#1-\pgflinewidth), inner sep=0pt, outer sep=0pt},
cross/.default={1pt}}

\newcommand{\om}{\omega}
\newcommand{\la}{\lambda}

\newcommand{\vare}{\varepsilon}

\newcommand{\bs}{\boldsymbol}
\newcommand{\noi}{\noindent}

\definecolor{my-purple}{RGB}{205,0,165}

\usepackage[normalem]{ulem}

\begin{document}
\title{Thin viscoelastic dewetting films of Jeffreys type subjected to gravity and substrate interactions}
\author{Valeria Barra\inst{1}\thanks{\emph{Email address:} \textrm{vb82@njit.edu}} \and Shahriar Afkhami\inst{1} \and Lou Kondic\inst{1}
%
}                     
%
%
\institute{\textsuperscript{1}Department of Mathematical Sciences, New Jersey Institute of Technology, Newark, NJ, 07102, USA}
\date{Received: date / Revised version: date}
%
\abstract{
This work presents a study of the interfacial dynamics of thin viscoelastic films subjected to the gravitational force and substrate interactions induced by the disjoining pressure, in two spatial dimensions. The governing equation is derived as a long-wave approximation of the Navier-Stokes equations for incompressible viscoelastic liquids under the effect of gravity, with the Jeffreys model for viscoelastic stresses. For the particular cases of horizontal or inverted planes, the linear stability analysis is performed to investigate the influence of the physical parameters involved on the growth rate and length scales of instabilities. Numerical simulations of the nonlinear regime of the dewetting process are presented for the particular case of an inverted plane. Both gravity and the disjoining pressure are found to affect not only the length scale of instabilities, but also the final configuration of dewetting, by favoring the formation of satellite droplets, that are suppressed by the slippage with the solid substrate.
%
} 

\authorrunning{Barra, Afkhami, Kondic}
\titlerunning{Thin viscoelastic films subjected to gravity and substrate interactions}
\maketitle
%

%

%
%

%

%


\section[Introduction]{Introduction}\label{sec:IntroInclinedPlane}

Wetting or dewetting phenomena are of great importance for different types of scientific and industrial processes, such as painting, coating or printing. For this reason, free-boundary or interfacial flows have been intensively studied (see, for instance, the reviews \cite{DeGennes2,ScardovelliZaleski} and references therein). To capture the interface instabilities, the position of the interface, or boundary between the different phases, needs to be modeled and found as a part of the solution of the equations governing the fluid flow \cite{TryggvasonEtAl}. The goal of the present work is to provide the mathematical description and a numerical investigation of the interfacial dynamics for free-boundary flows of thin layers of fluids, in the particular case in which the liquid of interest is a viscoelastic fluid of Jeffreys type \cite{Jeffreys}, subjected to both the disjoining pressure with the substrate \cite{Isreaelachvili} and the gravitational force.

Among the broad spectrum of natural or synthetic thin layers of liquids, viscoelastic films are ubiquitous. Polymeric liquids, in particular, are one example of viscoelastic liquids, constituted by a Newtonian (viscous) solvent and a non-Newtonian (polymeric) solute, and, possess interesting features, such as the stress relaxation and ``fading memory'' \cite{Bird}. As other suspensions or emulsions, polymeric liquids are considered one type of complex fluids. Viscoelastic liquids are, in turn, part of the wider class of non-Newtonian fluids. They are characterized by the fact that the stress tensor is dependent on the strain rate via a relationship that is generally nonlinear, and that can be, as in the model used in the present work, differential.

Historically, the foundations of the long-wave (or lubrication) theory were laid in a pioneering work by Reynolds \cite{Reynolds}, that analyzed the behavior of a viscous liquid confined between a solid substrate and a fluid-lubricated slipper bearing \cite{Oron}. Subsequently, the understanding of the interfacial instability phenomena arising in polymeric films dewetting substrates has motivated many theoretical and experimental studies, see, e.g.,~\cite{B-W,Reiter,Safran,GabrieleEtAl2}. For the case in which the gravitational body force is neglected, the long-wave formulation for thin viscoelastic films of Jeffreys type, subjected to substrate interactions, was developed by Raus\-cher et al.~\cite{Rauscher2005}, and subsequently treated in regimes of weak and strong slippage with the substrate by Blos\-sey et al.~\cite{Blossey2006}, as well as in the book \cite{BlosseyBook}. A numerical investigation of the formulation provided by Rauscher et al.~\cite{Rauscher2005} was performed by the authors in \cite{BarraEtAl}. Moreover, numerical studies concerning wetting and dewetting thin viscoelastic films, in the case of absence of gravity and slippage with the substrate, were carried out, for instance, by Tomar et al.~\cite{TomarEtAl}, and, more recently, by Benzaquen et al.~\cite{BenzaquenEtAl}, who have used the Maxwell model \cite{Maxwell} to describe viscoelastic stres\-ses. However, to the best of our knowledge, the derivation of a thin film equation for viscoelastic films of Jeffreys type under the effects of both the gravitational force and the disjoining pressure, that also includes the slippage with the substrate, is missing in the literature. Hence, such a derivation is proposed in the pre\-sent study.

Among the different types of interfacial instabilities, the ones due to thin films flowing down inclined planes have been widely studied for the case of viscous liquids (see, for instance,~\cite{Huppert,Hsieh,Schwartz,KellyGoussis,Hsieh2,Kondic2003,GombaKondic}). Most recently, dripping phenomena due to the competition between the Rayleigh-Taylor \cite{Kull} (RT) and Kapit\-za \cite{Kapitza} instabilities have been explored for viscous films flowing down planes, inclined at angles larger than $\pi/2$ with respect to the base, using the Weighted Residual Integral Boundary Layer model \cite{KofmanEtAl}. Moreover, in recent times, some industrial applications, such as the manufacturing of very thin (and possibly flexible) displays, have motivated similar investigations involving complex fluids, such as nematic liquid crystals \cite{MichaelLam}, vari\-able-viscosity fluids \cite{Tshehla}, or shear-thin\-ning liquids \cite{PicchiEtAl}. Moreover, Kaus and Becker have carried out a numerical investigation focusing on the RT instabilities of bi-layers of Maxwell fluids deposited on viscous ones \cite{KausBecker}, in which they confirmed that elasticity speeds up the growth of the RT instability. However, to the best of our knowledge, a similar numerical investigation that uses the Jeffreys viscoelastic model is not available in the literature. The goal of this study is not only to investigate the effects of viscoelasticity on the interfacial flows of thin viscoelastic films hanging on inverted planes, but also to analyze the competing effects of the different forces at play.

For thin film approximations, the van der Wa\-als attraction/repulsion interaction force is used to model the film breakup and the consequent dewetting process, as well as to impose the contact angle with the solid substrate. This force induces an equilibrium film on the solid substrate, leading to a prewetted (also called precursor) layer in nominally dry regions. We remark that in the absence of other forces, such as gravity, the liquid-solid interaction force is the only possible driving mechanism of dewetting. Moreover, we notice that the dynamics of dewetting processes can be divided in two regimes: the initial stage of the evolution, characterized by amplitudes in the interface thickness that are small relative to the initial height of the film and whose growth is analytically predicted by a Linear Stability Analysis (LSA); and the developed phase of the interface evolution, characterized by amplitudes that are no longer small and therefore cannot be described in terms of linear asymptotic approximations and require a fully nonlinear dynamic description.

In this work, we outline the theoretical and numerical study concerning the interfacial flow of two-dimensional thin viscoelastic films under the effects of both the gravitational force and the disjoining pressure. We derive here a novel governing equation describing the evolution of the interface of thin viscoelastic films lying on planes that can have an arbitrary inclination with respect to the base, obtained as a long-wave approximation of the Navier-Stokes equations, with the Jeffreys model for viscoelastic stresses. For the particular cases of horizontal or inverted planes, the LSA is performed to assess the effects of the different physical parameters involved on the dynamics governing the linear regime, and compare theoretical predictions of the early stage of the dynamics, with the numerical results obtained. The competing effects of the physical parameters involved on the length and time scales of instabilities are analyzed, in the linear and nonlinear regimes.

We find that, in the linear regime, the critical and most unstable wavenumbers are neither dependent on the viscoelastic parameters, nor on the slip length, but only on the interactions induced by the disjoining pressure and the gravitational force. Moreover, we provide numerical simulations of the evolution of the interface in the nonlinear regime, for the particular case of an inverted plane. In this regime, we find that the gravitational force and the disjoining pressure affect the equilibrium configuration attained by the dewetted films, by favoring the formation of satellite droplets, and by forming a hump in the nominally dry central region, destabilizing the precursor film. Moreover, we notice that the slippage with the solid substrate suppresses the secondary droplets, observed in regimes of microgravity.

We emphasize that the Jeffreys model (together with the Maxwell model \cite{Maxwell} and the Newton model for viscous fluids \cite{Bird}) is appropriate to describe liquids in which displacement gradients are small \cite{Bird,BarraEtAl2}. More broadly, generalizations of the Jeffreys model, that include nonlinear objective stress rates (in place of the simple time derivative of the stress) to describe the advection and rotation of the stress tensor with the flow, are considered; for instance, the Oldroyd-B model \cite{Bird} that uses an upper-convected time derivative, or the corotational Jeffreys model \cite{MunchWagner} that involves the Jaumann derivative. However, in the context of the present work, and similarly in studies concerning thin viscoelastic dewetting films in the absence of gravity (see, e.g.,~\cite{Rauscher2005,BarraEtAl,TomarEtAl}), the liquid films are considered to be initially at rest and undergo a spontaneous, relatively slow, dewetting flow. In \cite{BarraEtAl,TomarEtAl}, it has been found that viscoelastic dewetting films exhibit a slow, viscous response, especially in the early times and final stages of evolution. As Tomar et al. have discussed in \cite{TomarEtAl}, results obtained from nonlinear simulations that use linear viscoelastic models without the upper-con\-vected terms are expected to be qualitatively accurate. In particular, this applies to the regions of the decay of the capillary ridge (as also discussed by Rauscher et al.~in \cite{Rauscher2005}), and in the central hole where secondary length scales of instabilities form and slowly coalesce (that is the region where most of the attention is focused on in the present work). Although limited, both the Maxwell and the Jeffreys models have been applied to and proven to be useful for the analysis of a broad range of materials (see, e.g.,~\cite{Larson}), and analyses using these models have served as baselines for studies that have considered their expansions including nonlinearities.

The remainder of this paper is organized as follows. In \S~\ref{sec:MathematicalFormulationInclinedPlane}, we outline the mathematical modeling, whose detailed description and thin film approximation are given in the Appendix. In \S~\ref{sec:InclinedThinFilmsNumerics}, we outline the numerical methods employed. In \S~\ref{sec:InclinedThinFilmsResults}, we discuss our numerical results for the linear and nonlinear regimes, and finally, in \S~\ref{sec:InclinedThinFilmsConclusions}, we draw our conclusions.

\section[Mathematical Formulation]{Mathematical Formulation}\label{sec:MathematicalFormulationInclinedPlane}

\begin{figure}[t]
\centering
{\includegraphics[width=.95\linewidth]{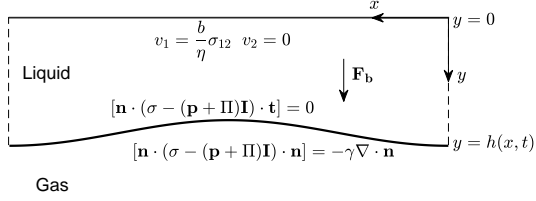}}
\caption{Schematic of a fluid interface and boundary conditions at the interface of the fluid and the solid substrate, for $\alpha = \pi$ (inverted plane).}\label{fig:InvertedPlaneSetup}
\end{figure}

We present here the governing equation (whose detailed derivation is shown in the Appendix) for a viscoelastic fluid lying on a plane, inclined at an angle $\alpha$ with respect to the positive $x$-axis, and subjected to the disjoining pressure and weak slip regime with the substrate. We consider an incompressible liquid, with constant density $\rho$, surrounded by a gas phase assumed to be inviscid, dynamically passive, and of constant pressure. The equations of conservation of momentum and mass, respectively, for the liquid phase are
\begin{subequations}\label{Eq:N-SChapter3}
\begin{align}
\rho \left( \partial_t \textbf{v} + \textbf{v} \cdot \nabla \textbf{v}\right) &= - \nabla (p + \Pi) + \nabla \cdot \bs{ \sigma} + \mathbf{F_b} \ , \label{GovMomentumChapter3}\\
\nabla \cdot \textbf{v} &= 0 \, , \label{IncompressibilityChapter3}
\end{align}
\end{subequations}
\noindent where $\textbf{v} = (v_1(x,y,t),v_2(x,y,t))$ is the velocity field in the Cartesian $xy$-plane (as per convention, the $x$-axis is parallel to the plane, and the $y$-axis is perpendicular to the plane), and $\nabla= (\partial_x, \partial_y)$; $\bs{ \sigma}$ is the (symmetric) stress tensor, $p$ is the pressure, $\Pi = \Pi (h)$ is the disjoining pressure due to liquid-solid interaction forces (we note that $\nabla \Pi =0$ except at the liquid-gas interface), and $\mathbf{F_b} = (\rho g \sin \alpha, - \rho g \cos \alpha)$, where, the gravitational acceleration constant is positive ($g>0$) for the reference system depicted in Figure \ref{fig:InvertedPlaneSetup}.

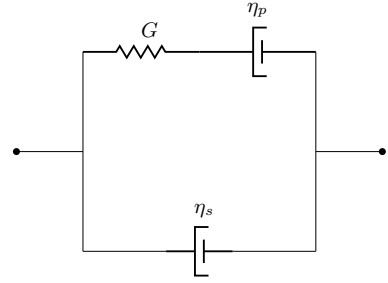
\begin{figure}[t]
\centering
\resizebox{.7\linewidth}{!}{
\begin{tikzpicture}
\tikzstyle{spring}=[thick,decorate,decoration={zigzag,pre length=0.5cm,post length=0.5cm,segment length=6}]
\tikzstyle{damper}=[thick,decoration={markings,
  mark connection node=dmp,
  mark=at position 0.5 with
  {
    \node (dmp) [thick,inner sep=0pt,transform shape,rotate=-90,minimum width=20pt,minimum height=3pt,draw=none] {};
    \draw [thick] ($(dmp.north east)+(2pt,0)$) -- (dmp.south east) -- (dmp.south west) -- ($(dmp.north west)+(2pt,0)$);
    \draw [thick] ($(dmp.north)+(0,-5pt)$) -- ($(dmp.north)+(0,5pt)$);
  }
}, decorate]

\draw [spring](0,3) -- (1.75,3) ;
\draw (1,3) node[anchor=south,yshift=3pt] {$G$};
\draw [damper](1.75,3) -- (3.5,3) ;
\draw (2.52,3.3) node[anchor=south,yshift=3pt,xshift=3pt]{$\eta_p $};

\draw (0,0) -- (0,3);
\draw (3.5,0) -- (3.5,3);

\draw (0,0) -- (1.25,0);
\draw (2.25,0) -- (3.5,0);

\draw [damper](1.25,0) -- (2.25,0) ;
\draw (1.72,.3) node[anchor=south,yshift=3pt,xshift=3pt] {$\eta_s $};
\draw (3.5,1.5) -- (4.5,1.5);
\fill[black] (4.5,1.5) circle(1.5pt);
\draw (0,1.5) -- (-1,1.5);
\fill[black] (-1,1.5) circle(1.5pt);


\end{tikzpicture}
}
\caption{Jeffreys model represented as a mechanical system, where $G$ represents the shear modulus, and $\eta_s$ and $\eta_p$ the viscosity coefficients of the Newtonian solvent and the polymeric solute, respectively.}\label{fig:JeffreysScheme}
\end{figure}

To model the stresses, we use a generalization of the Maxwell model \cite{Maxwell} for viscoelastic liquids: the Jeffreys model \cite{Jeffreys}. The Jeffreys model, together with the Maxwell model, belongs to a class of linear viscoelastic differential models. Viscoelastic materials can be described as mechanical systems where material points are connected by dashpots (representing energy dissipating devices), springs (representing energy storing devices), or any combination of the two devices. For a Maxwell material \cite{Maxwell}, the system comprises a spring and a dashpot in series, and for a Kelvin-Voigt material \cite{Siginer}, a spring and a dashpot in parallel. For a Jeffreys fluid, the system is composed of another dashpot connected in parallel to a Maxwell system \cite{Siginer,MainardiSpada,Gutierrez-Lemini} (see Figure \ref{fig:JeffreysScheme}).

The constitutive law for the Jeffreys model is given by
\begin{equation}
\label{Jeffreys}
\bs{\sigma} + \lambda_1 \partial_t \bs{\sigma} = 2 \eta (\dot{\bs{\epsilon}} + \lambda_2 \partial_t \dot{\bs{\epsilon}}) \, ,
\end{equation}
\noi where $\dot {\boldsymbol{\epsilon}}$ is the strain rate tensor, e.g.~$\dot{\epsilon}_{ij}=  (\partial v_j / \partial {x_i} + \partial v_i / \partial {x_j})/2$, with $i,j=\{1,2\}$, and $\eta$ is the dynamic viscosity coefficient. In this model, the response to the deformation of a viscoelastic liquid is characterized by two time constants, $\la_1$ and $\la_2$, the \textit{relaxation time} and the \textit{retardation time}, respectively, related by $\lambda_2 = \la_1 \eta_s/(\eta_s + \eta_p)$. Here, $\eta_s$ and $\eta_p$ are the viscosity coefficients of the Newtonian solvent and the polymeric solute, respectively, such that $\eta = \eta_s + \eta_p$ (see Figure \ref{fig:JeffreysScheme}). Noting that the ratio $\eta_s/(\eta_s + \eta_p) \leq 1$, we have that $\la_1 \geq \la_2$ \cite{Bird}. We observe that the Maxwell model \cite{Maxwell} is recovered from the Jeffreys model when $\la_2=0$, and the Newtonian constitutive relation for viscous fluids \cite{Bird} is obtained when $\la_1=\la_2$. We emphasize that it is believed to be quite restrictive to expect a polymeric liquid of a broad molecular weight distribution to be characterized in terms of a single relaxation time \cite{Siginer}. In fact, the relaxation time occurring in the Jeffreys (and Maxwell) constitutive relation is interpreted as the longest relaxation time exhibited by a polymeric liquid \cite{TomarEtAl}. In the present work, the Jeffreys model is preferred over the simpler Maxwell model to be able to describe features such as the retardation due to the Newtonian solvent in the solution. In \S~\ref{sec:InclinedThinFilmsResults}, we will analyze the influence of the relaxation and retardation time constants, as well as the other physical parameters, on the dynamics and morphologies of the dewetting films.

The system of equations (\ref{Eq:N-SChapter3}) is subjected to boundary conditions at the free surface, represented parametrically by the function $f(x,y,t)=y-h(x,t)=0$, and boundary conditions at the solid substrate ($y=0$). At the latter, we apply the non-penetration and the Navier slip boundary conditions, with the slip length coefficient denoted by $b \geq 0$. As also discussed in \cite{Fetzer,Munch}, long-wave models for thin films can be derived in different slip regimes. In this work, we will focus on the weak slip regime.

The stress balance at the interface is given by
\begin{equation}\label{Eq:LaplacePressBCChapter3}
\left( \bs{\sigma} - (p+\Pi) \mathbf{I} \right) \cdot \mathbf{n} = \gamma \kappa \mathbf{n} \, ,
\end{equation}
\noindent where $\mathbf{I}$ is the identity matrix. In the absence of motion, this condition describes the jump in the pressure across the interface with outward unit normal $\mathbf{n}$ (whose definition is given in the Appendix), and a local curvature, $ \kappa= - \nabla \cdot \mathbf{n}$, due to the surface tension $\gamma$. The form of the disjoining pressure, $\Pi = \Pi(h)$, used in this work is given by the following expression
\begin{align}\label{dimensionalVdW}
\Pi = \frac{\gamma(1 - \cos \theta_e)}{M h_{\star}} \left[ {\left( \frac{h_{\star}}{h}  \right)}^{m_1} - {\left( \frac{h_{\star}}{h}\right)}^{m_2} \right] \, ,
\end{align}
\noi where $h_{\star}$ represents the equilibrium film thickness induced by the van der Wa\-als attraction/repulsion interaction force, $\theta_e$ is the equilibrium contact angle, formed between the fluid interface $y=h(x,t)$ and the solid substrate, and $M = (m_1-m_2)/[(m_2-1)(m_1-1)]$, where $m_1$ and $m_2$ are constants such that $m_1>m_2>1$. In this work, we choose $m_1=3$ and $m_2=2$, as also widely used in the literature, for instance by the authors in \cite{DiezKondic2007,Teletzke,Ivana}, but other values can be modeled as verified in \cite{Kyle}.

The interested reader can find the derivation of the governing equation for the evolving interface $h(x,t)$ in the Appendix, where all quantities and scalings are defined. We report here its final dimensionless form in which we use the compact notation $\partial (\cdot)/\partial x = (\cdot)_x$, and $\partial (\cdot)/\partial t = (\cdot)_t$, (similarly for higher order spatial or temporal derivatives later on in the text)
\begin{align}\label{Eq:GovEqnWithGravity1D}
& (1 + \lambda_2 \partial_t) h_t + (\lambda_2 -\lambda_1)  \left[ \left( \frac{h^2}{2} {Q} - h {R}  \right)  h_t \right]_x + \nonumber \\
& \left[ (1 + \lambda_1 \partial_t ) \frac{h^3}{3} \left(-p_x  +\mathcal{S}\right) + \right.\nonumber \\
&\left. (1 + \lambda_2 \partial_t ) bh^2 \left( -p_x  + \mathcal{S} \right)   \right]_x = 0 \, ,
\end{align}
\noindent in which we use the dimensionless form of the pressure, $p$, given in the Appendix, and where $Q$ and $R$ satisfy the equations:
\begin{subequations}\label{Eq:Q&RWithGravity1D}
\begin{align}
(1 + \la_2 \partial_t) {Q} =& p_x - \mathcal{S}  \, , \label{Eq:QWithGravity1D}\\
(1 + \la_2 \partial_t) {R} =& h\left( p_x - \mathcal{S} \right)  \, . \label{Eq:RWithGravity1D}
\end{align}
\end{subequations}
\noindent In equations (\ref{Eq:GovEqnWithGravity1D}) and (\ref{Eq:Q&RWithGravity1D}), we have used
\begin{subequations}\label{Def:S&C}
\begin{align}
\mathcal{S} &= \frac{\rho g L^2 \vare^2}{V \eta} \sin \alpha \, \\
\mathcal{C} &= \frac{\rho g L^2 \vare^3}{V \eta} \cos \alpha \, .
\end{align}
\end{subequations}
\noindent The reader can find the meaning of the scaling factors $L,\Gamma,V$, together with the definition of the small parameter $\vare$, in the Appendix. Let $ \mathcal{B} = \rho g L^2/ \Gamma \equiv \rho g L^2 \vare^3 / V \eta = O(1)$ (unless specified differently) be the Bond number, a dimensionless quantity representing the importance of gravity relative to surface tension. We note that for the particular cases in which the plane has small inclination $\alpha$ with the base, i.e., for $\alpha = \vare \alpha^*$ or $\alpha = \pi + \vare \alpha^*$, the parameters in equations (\ref{Def:S&C}) can be expressed as $\mathcal{S} \approx \mathcal{B} \alpha^*$ and $\mathcal{C} \approx \mathcal{B}$, respectively.

We notice that for the case in which $\la_1 = \la_2$ (that corresponds to a Newtonian fluid), equations (\ref{Eq:GovEqnWithGravity1D}) and (\ref{Eq:Q&RWithGravity1D}) reduce to the governing equation for thin viscous films flowing down inclined planes, as outlined in \cite{Kondic2003,TeSheng}. Moreover, we remark that, for the particular cases for which we present our results in this work (i.e., for $\alpha = 0,\pi$), the term $\mathcal{S}$ cancels. However, it is retained from now on, to allow for easy extensions of the present study in future research endeavors, that may consider arbitrary values of the inclination angle $\alpha$.

\section[Numerical Methods]{Numerical Methods}\label{sec:InclinedThinFilmsNumerics}

To discretize equation (\ref{Eq:GovEqnWithGravity1D}), we isolate the time derivatives from the spatial ones, so that we can apply an iterative scheme to find the approximation to the solution at the new time step. We do so by differentiating the spatial derivatives and, assuming the partial derivatives of $h(x,t)$ to be continuous, obtaining
\begin{align}\label{Eq:ThinViscoelasticIncline}
& \la_2h_{tt} + { \left[\left( \frac{h^3}{3}   + bh^2 \right)\left(-p_x  +\mathcal{S}\right) \right]_x }+ \nonumber \\
&{ \left\{1 + (\la_2 - \la_1 ) \left[ \left( \frac{h^2}{2} {Q} - h {R} \right)_x \right]\right\}} h_t  +\nonumber \\
&{\left( h_x\right)_t} {(\la_2 - \la_1) \left( \frac{h^2}{2} {Q} - h {R}  \right)} + \nonumber \\
&\la_1  {\partial_t \left[  \left( \frac{h^3}{3}\left(-p_x  +\mathcal{S}\right) \right)_x \right] } + \nonumber \\
&\la_2  { \partial_t \left[ \left(bh^2 \left(-p_x  +\mathcal{S}\right)  \right)_x \right] } =0 \, .
\end{align}
\noi The spatial domain $[0, \Lambda]$ is discretized by a staggered structured grid, in which the first and third order derivatives are defined at the cell-centers, and the second and fourth order ones at the grid points. Following the natural order from left to right, adjacent vertices are associated to the indices $i-1,i,i+1$, respectively. Thus, we let $x_{i}=x_0+i \Delta x ,\; i=1,2, \ldots, \, N$ (where $N=\Lambda / \Delta x$, and $\Delta x$ is the fixed grid size), so that the endpoints of the physical domain, $0$ and $\Lambda$, correspond to the $x_{1}- \frac{\Delta x}{2}$ and $x_{N}+\frac{\Delta x}{2}$ cell-centers, respectively. Similarly, we discretize the time domain and denote by $h_i^n$ the approximation to the solution at the point ($x_i, n\Delta t$), where $n=0,1, \ldots$ indicates the number of time steps, and $\Delta t$ is the temporal step size, which can be chosen adaptively to speed up the marching algorithm when the solution does not exhibit fast temporal variations (see \cite{BarraEtAl,DiezKondic2007,DiezKondic2002} for a detailed description). In addition to the numerical formulation provided in \cite{BarraEtAl}, the discrete versions of equations (\ref{Eq:QWithGravity1D}) and (\ref{Eq:RWithGravity1D}) are given by
\begin{subequations}\label{Eq:DiscreteQ&RWithGravity}
\begin{align}
\frac{{Q}^{n+1}_i - {Q}^n_i}{\Delta t} &= - \frac{{Q}_i^n}{\la_2} - \frac{1}{\la_2} \left( -p_x + \mathcal{S} \right)_i^n \, , \label{Eq:DiscretizedQ} \\
\frac{{R}^{n+1}_i - {R}^n_i}{\Delta t} &= - \frac{{R}_i^n}{\la_2} - \frac{1}{\la_2} h_i^n \left( -p_x  + \mathcal{S} \right)_i^n  \, ,  \label{Eq:DiscretizedR}
\end{align}
\end{subequations}
\noindent that we solve using the forward Euler method with initial conditions ${Q}_i^0 = 0$ and ${R}_i^0 = 0$, respectively. The nonlinear terms $h^2$ and $h^3$ are computed at the cell-centers, as outlined in \cite{BarraEtAl,Kondic2003,Bertozzi}.

We solve equations (\ref{Eq:ThinViscoelasticIncline}) and (\ref{Eq:DiscreteQ&RWithGravity}) for the particular case in which $\alpha = \pi$ (according to the setup depicted in Figure \ref{fig:InvertedPlaneSetup}), and at the endpoints of the domain, we impose the $h_x=h_{xxx}=0$ boundary conditions, that reflect the symmetry of the problem. The condition $h_x=0$ gives the value of $h$ at the two ghost points $x_{0}$ and $x_{N+1}$ outside the physical domain, i.e.~$h_{0}=h_{1}$ and $h_{N+1}=h_{N}$; the condition $h_{xxx}=0$ specifies the two ghost points $x_{-1}$ and $x_{N+2}$, i.e.~$h_{-1}=h_2$ and $h_{N+2}=h_{N-1}$ (consistent with \cite{Kondic2003}). Moreover, similarly to the numerical study in \cite{BarraEtAl}, we apply a mixed implicit/explicit finite difference formulation to discretize the nonlinear equation (\ref{Eq:ThinViscoelasticIncline}). We remark that, because of the symmetry condition, we can reduce the computational domain to half the physical domain. However, in the results that follow, we show the numerical solutions on the entire domain for ease of visualization.

After Newton's linearization, we obtain a system of equations of the form $A\xi=B$, that we numerically solve for the correction term, $\xi$, using a direct method \cite{TeSheng}. The initial condition given for $h(x,0)$ is a known function that describes the initial perturbation of the fluid interface (see \S~\ref{sec:InclinedThinFilmsResults}), and the initial velocity is $h_t (x,0) = 0$, describing the fact that the considered films are initially at rest.

\section[Results and Discussion]{Results and Discussion}\label{sec:InclinedThinFilmsResults}

\subsection[Linear Stability Analysis]{Linear Stability Analysis}\label{LSAInclinedPlane}

In this section, we report our results regarding the special case of an inverted plane, for which $\alpha = \pi$, as depicted in Figure \ref{fig:InvertedPlaneSetup}. We perturb a flat film of initial thickness $h_0$ by an oscillatory Fourier mode of amplitude $\delta h_0$ (such that $\delta \ll 1$), with wavenumber $k$, i.e., we let $h(x,t)= h_0 + \delta h_0 e^{ikx + \omega t}$. Performing the LSA on equations (\ref{Eq:GovEqnWithGravity1D}) and (\ref{Eq:Q&RWithGravity1D}) leads to the following dispersion relation
\begin{align}\label{Eq:CompleteLSAWithGravity}
&\la_2 \om^2 + \left[1 + (\mathcal{K_C} +i\mathcal{S}k )\left( \la_1 \frac{h_0^3}{3} +\la_2 bh_0^2 \right)\right]\om +\nonumber \\
&(\mathcal{K_C} +i\mathcal{S}k )\left( \frac{h_0^3}{3} + bh_0^2 \right)=0 \, ,
\end{align}
\noindent where we have defined
\begin{align}
\mathcal{K_C} \coloneqq k^4 - k^2 (\Pi'(h_0) - \mathcal{C} ) \, .
\end{align}
\noindent We consider the real part of the two roots of the dispersion relation (\ref{Eq:CompleteLSAWithGravity}), namely $Re\{\om_1\}$ and $Re\{\om_2\}$. One is always negative (indicating stable modes), say $Re\{\om_2\}$, and the other one has varying sign (describing potentially unstable ones), say $Re\{\om_1\}$. We find
\begin{align}\label{LSASolution}
&Re\{\om_{1}\} = \frac{-\left[1 + \mathcal{K_C} \left( \la_1 \frac{h_0^3}{3} +\la_2 bh_0^2 \right)\right]}{2\la_2} +\nonumber \\
&\frac{Re\{\sqrt{\Delta_{\om} } \}}{2\la_2} \, ,
\end{align}
\noi where we have defined
\begin{align}
&\Delta_{\om} \coloneqq {\left[1 + (\mathcal{K_C} +i\mathcal{S}k ) \left( \la_1 \frac{h_0^3}{3} +\la_2 bh_0^2 \right)\right]}^2 -\nonumber \\
&4 \la_2 (\mathcal{K_C} +i\mathcal{S}k)  \left( \frac{h_0^3}{3} + b h_0^2 \right) \, .
\end{align}
\noindent The critical wavenumber, for which $Re\{\om_{1}\} = 0$, satisfies the relationship $k_c^2 =\Pi'(h_0) - \mathcal{C}$. Hence, through this relationship, we note that the gravitational term affects the length scales of instability. We also remark that, for the parameters considered in this work, $\Pi'(h_0)>0$ and $- \mathcal{C}>0$. Moreover, we notice that the wavenumber of maximum growth, as for the case without gravity \cite{Rauscher2005,BarraEtAl}, satisfies the relationship $k_m =k_c / \sqrt{2}$. In addition, consistent with the previous studies without gravity \cite{BarraEtAl,TomarEtAl}, we find that the maximum growth rate, $\om_m=\om(k_m)$, is an increasing function of $\la_1$ and $b$, while a decreasing function of $\la_2$.

\subsection[Dewetting of Thin Viscoelastic Films on Inverted Substrates]{Dewetting of Thin Viscoelastic Films on Inverted Substrates}\label{sec:ThinFilmsResultsInclinedPlane}

\begin{figure}[t]
\centering
{\includegraphics[width=.9\linewidth,trim=.05cm 0.05cm .05cm 0.05cm,clip=true]{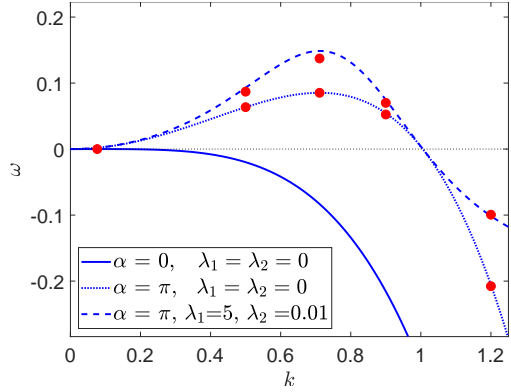}}
\caption{Comparison of the computed growth rates, corresponding to different wavenumbers (red dots), with the prediction of the LSA, for the parameters given in (\ref{parameters}), and, in particular, for a Newtonian film, i.e., with $\la_1=\la_2=0$, and $\alpha = \pi$ (blue dotted line) or $\alpha = 0$ (blue solid line); and for a viscoelastic film with $\la_1=5$ and $\la_2=0.01$, and $\alpha = \pi$ (blue dashed line).}\label{fig:LSAWithGravity}
\end{figure}

\begin{figure}[t]
\captionsetup{type=figure}
\centering
\subfloat[]{\includegraphics[width=.9\linewidth,trim=.05cm 0.05cm .05cm 0.0cm,clip=true]{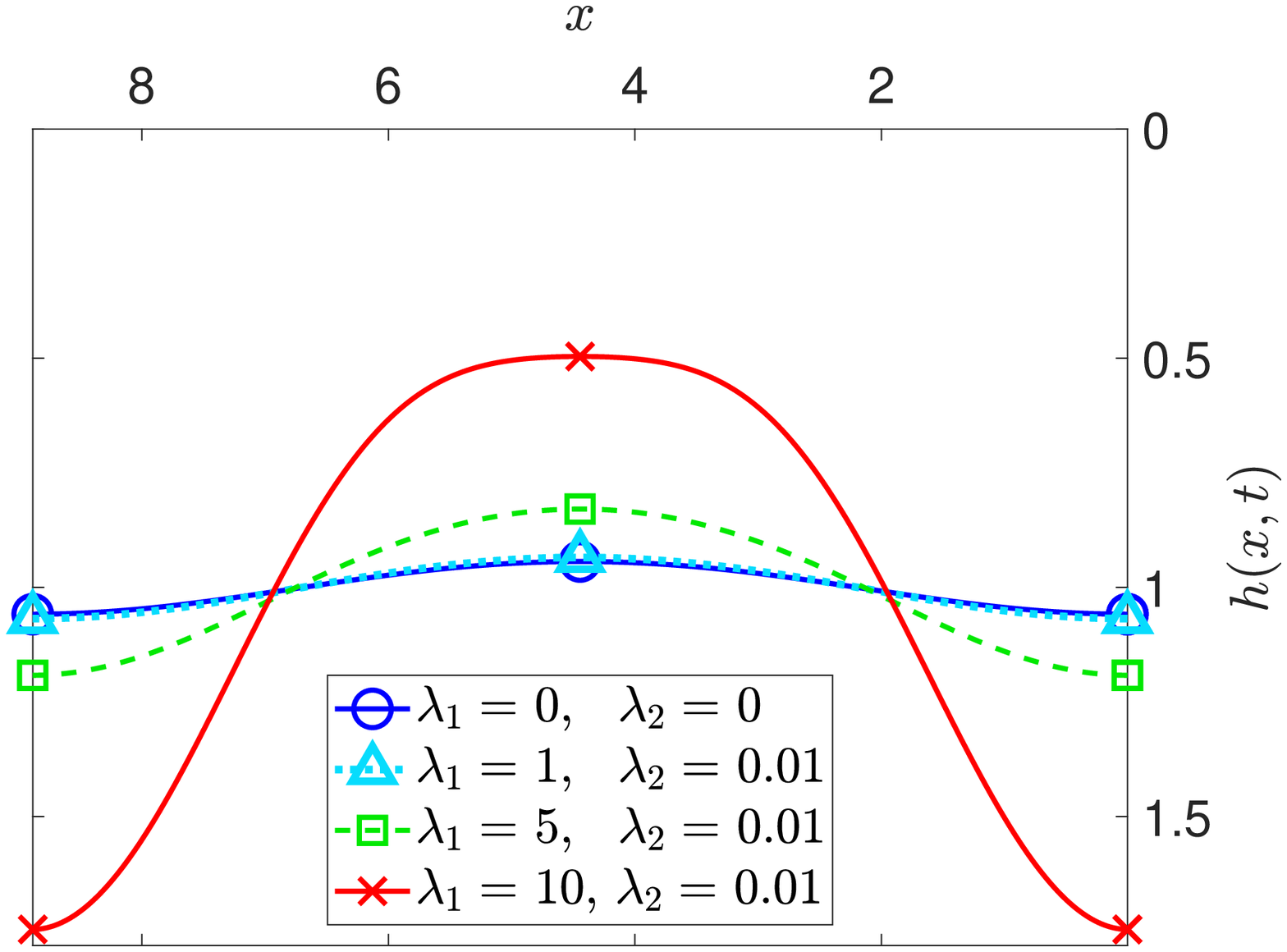}\label{fig:InvertedPlaneNoVdWa}}\\
\subfloat[]{\includegraphics[width=.9\linewidth,trim=.05cm 0.05cm .05cm 0.0cm,clip=true]{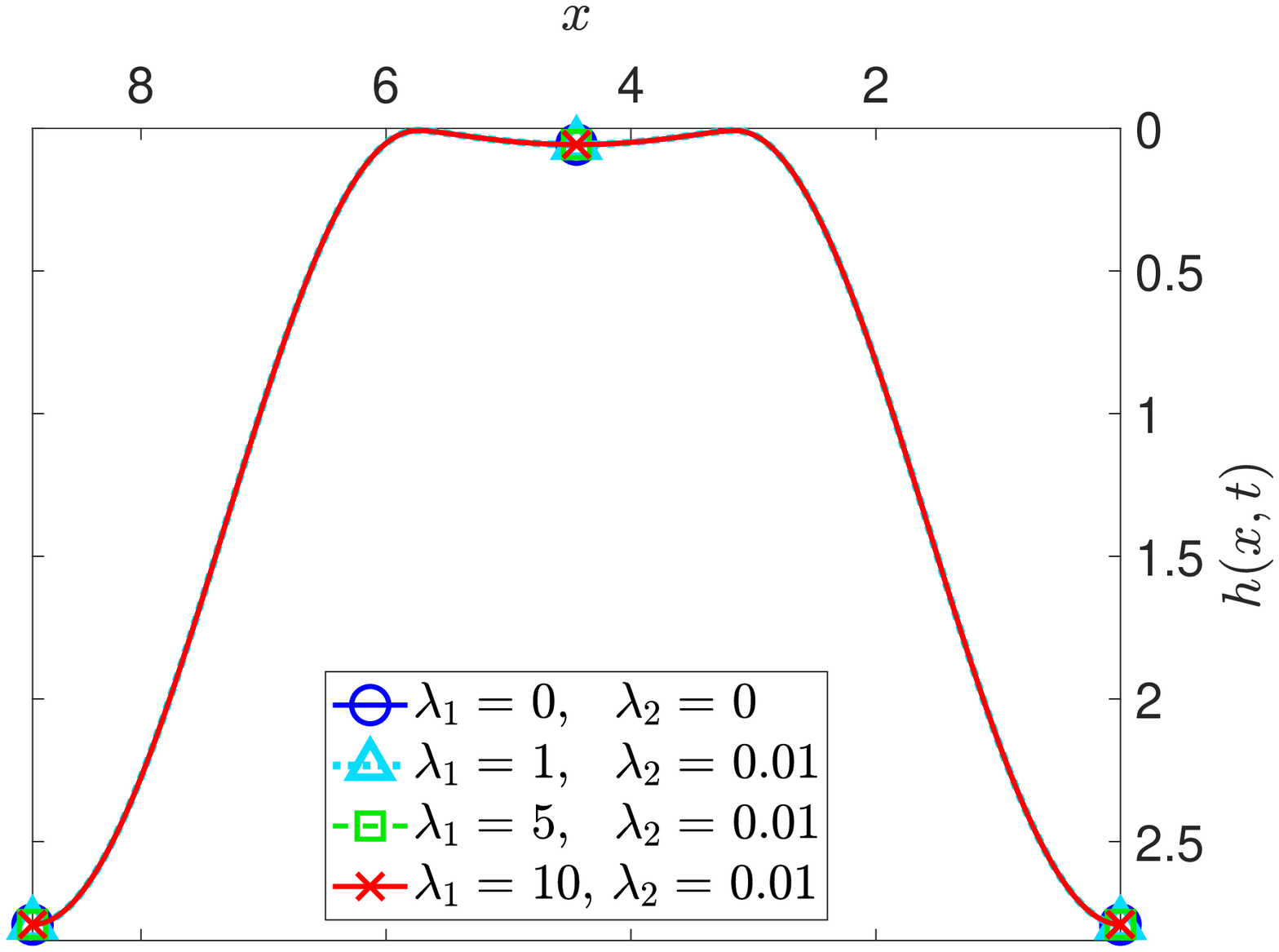}\label{fig:InvertedPlaneNoVdWb}}\\
\caption{Evolution of different viscoelastic films with the same parameters as in (\ref{parameters}), in the absence of the disjoining pressure, and: $\la_1 = \la_2 = 0$ (blue solid curve with circles), $\la_2 = 0.01$ and $\la_1 = 1$ (cyan dotted curve with triangles), $\la_1=5$ (green dashed curve with squares), and $\la_1=10$ (red solid curve with crosses), respectively; in \protect\subref{fig:InvertedPlaneNoVdWa}, at time $t=20$, and in \protect\subref{fig:InvertedPlaneNoVdWb}, at time $t=10^4$.}\label{fig:InvertedPlaneNoVdW}
\end{figure}

We outline here the numerical results for dewetting thin films under the influence of the disjoining pressure and the gravitational force. As anticipated in \S~\ref{LSAInclinedPlane}, we consider the particular case in which $\alpha=\pi$, that is, for films that hang on an inverted plane. As described in \S~\ref{LSAInclinedPlane}, we perturb the initially flat fluid interface of thickness $h_0$, with a perturbation characterized by the wavenumber $k=k_m$ and $\delta=0.01$, i.e., $h_0(x,0) = h_0  + \delta h_0  \cos (x k_m)$, and we choose the domain size, $\Lambda$, to be equal to the wavelength of maximum growth, that is, $\Lambda_m= 2 \pi / k_m$. Initially, we consider dewetting films in a regime of no-slip with the solid substrate, and subsequently, we analyze the effects of the substrate slippage on the dewetting dynamics and morphologies. For all the results that follow, we use a fixed grid size of $\Delta x = 0.01$; however, all results have been verified to be mesh independent.

\begin{figure}[t]
\captionsetup{type=figure}
\centering
\subfloat[]{\includegraphics[width=6cm,trim=.0cm 0.05cm .15cm 0.0cm,clip=true]{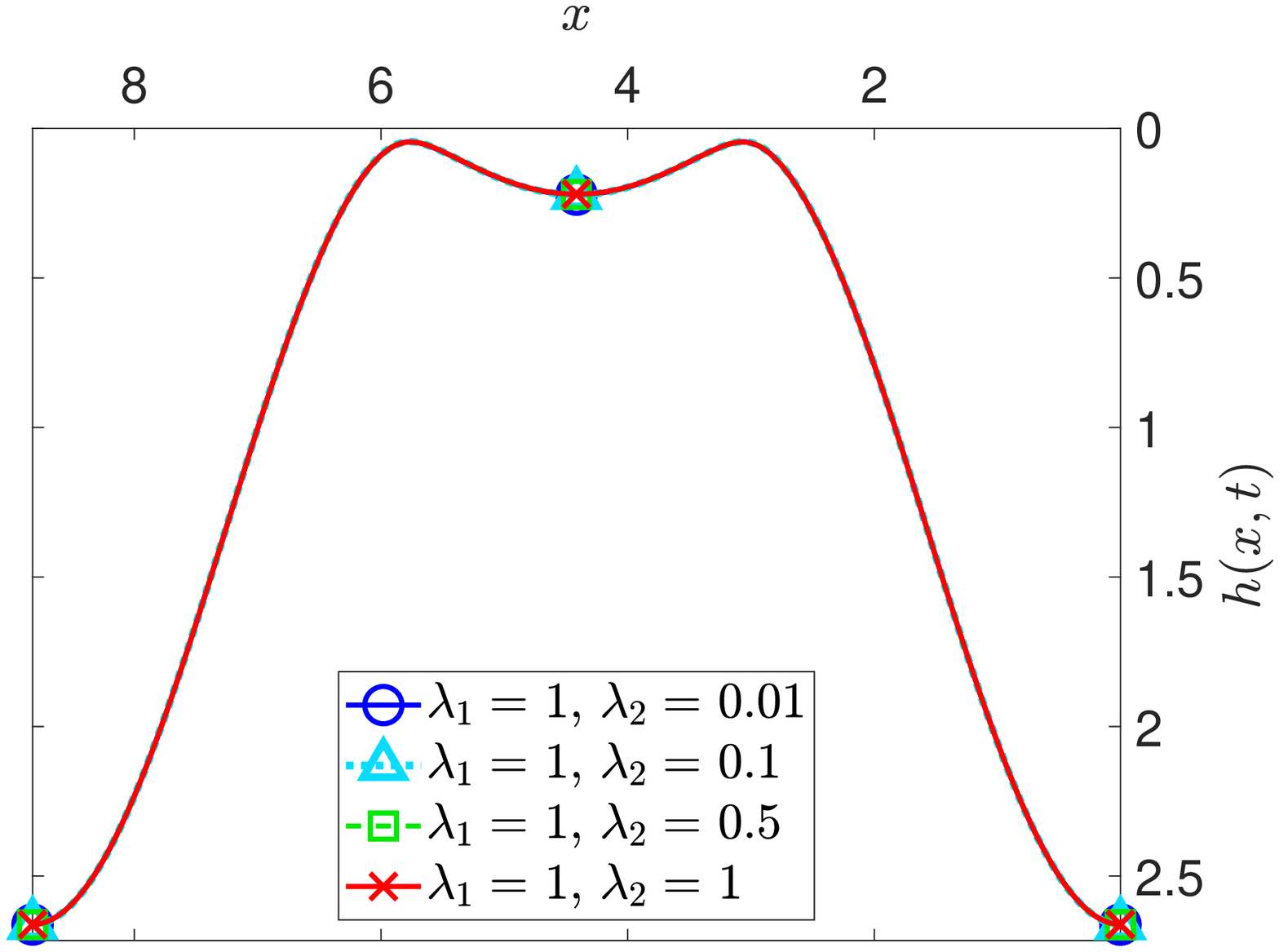}\label{fig:varyingL2a}}\\
\subfloat[]{\includegraphics[width=6cm,trim=.0cm 0.05cm .15cm 0.0cm,clip=true]{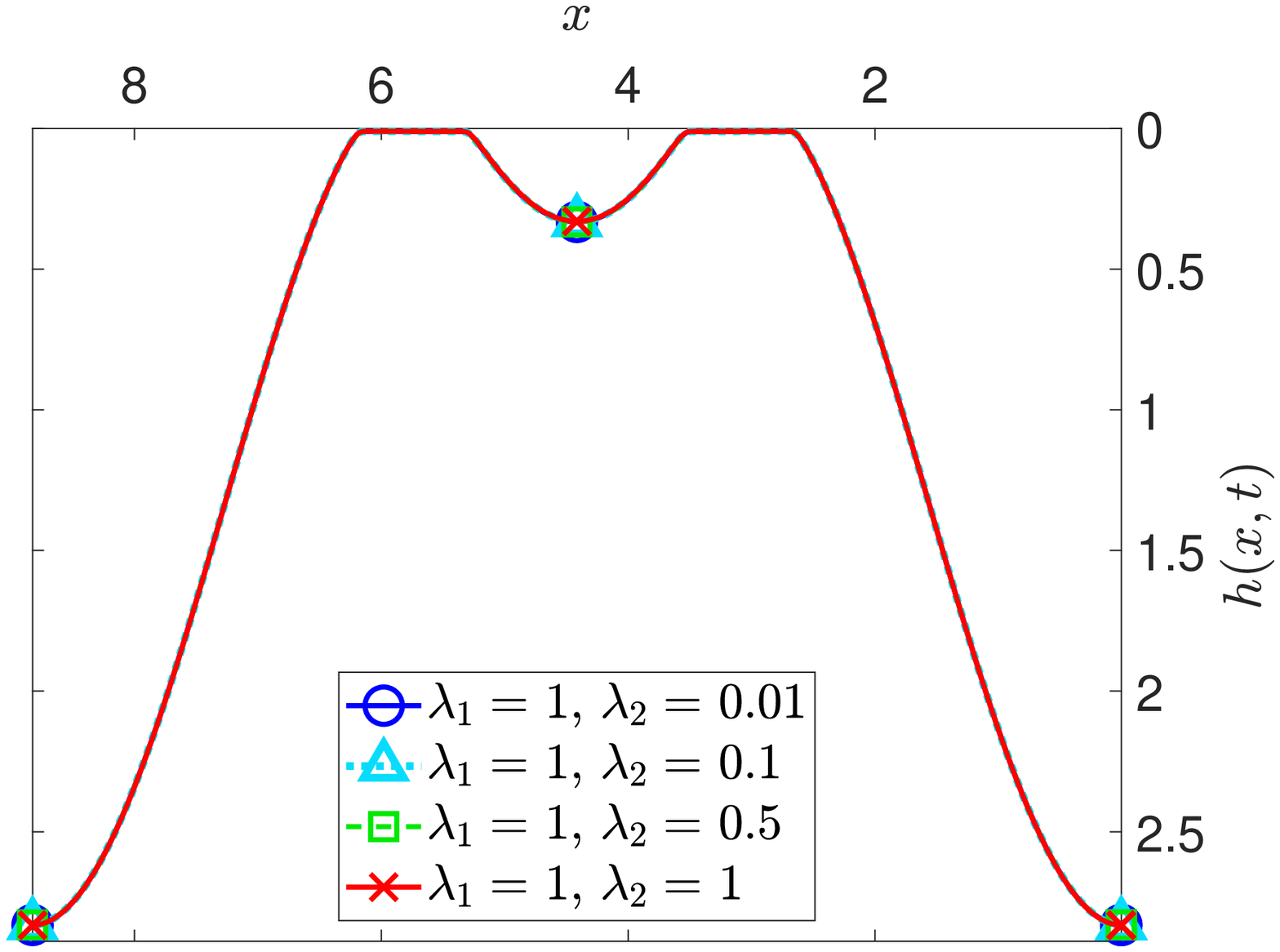}\label{fig:varyingL2b}}\\
\caption{Evolution of different viscoelastic films with the same parameters as in (\ref{parameters}), in the absence of the disjoining pressure, with $\la_1=1$ and $\la_2 = 0.01$ (blue solid curve with circles), $\la_2=0.1$ (cyan dotted curve with triangles), $\la_2=0.5$ (green dashed curve with squares), and $\la_2=1$ (red solid curve with crosses), respectively; in \protect\subref{fig:varyingL2a}, at time $t=1.47\times 10^2$, and in \protect\subref{fig:varyingL2b}, at time $t=10^4$.}\label{fig:varyingL2}
\end{figure}

To isolate the effects of the gravitational force, we start by analyzing the behavior of dewetting films for the particular case of the absence of the disjoining pressure with the substrate. In Figure \ref{fig:LSAWithGravity}, we present the comparison of the computed growth rates, corresponding to different wavenumbers (red dots), with the theoretical values predicted by the dispersion relation, given by equation (\ref{LSASolution}), for the following parameters
\begin{align}\label{parameters}
h_0=1, \; h_{\star}=0.01, \; \theta_e=\pi/4,\; \mathcal{B}=1, \; b=0 .
\end{align}
\noindent In Figure \ref{fig:LSAWithGravity}, in particular, we show the results for a Newtonian film, i.e., with $\la_1=\la_2=0$, for $\alpha = \pi$ (blue dotted line) or $\alpha = 0$ (blue solid line), and a viscoelastic film with $\la_1=5$ and $\la_2=0.01$, and $\alpha = \pi$ (blue dashed line). We remark that in this case, the only driving force for the instability is gravity. As expected, in the absence of the attraction/repulsion force with the substrate, and for $\alpha = 0$, there is no instability, and the growth rate is negative for all wavenumbers. To compare with the analytical values, we measure the computed growth rates in the linear regime in which the amplitude of the interface function grows exponentially.

In what follows, we will show numerical results for films dewetting an inverted substrate (i.e. for $\alpha = \pi$), with the same fixed initial and equilibrium thicknesses, equilibrium contact angle, and slip coefficient as shown in (\ref{parameters}), unless specified differently, and will vary the different physical parameters to investigate their isolated effects. In Figure \ref{fig:InvertedPlaneNoVdW}, we investigate the dynamics of thin viscoelastic films, in the absence of the disjoining pressure. We compare the evolution of a Newtonian film, with $\la_1 = \la_2 = 0$ (blue solid curve with circles), with the one of viscoelastic films, with $\la_2 = 0.01$ and $\la_1 = 1$ (cyan dotted curve with triangles), $\la_1 =5$ (green dashed curve with squares), and $\la_1 =10$ (red solid curve with crosses). Figure \subref*{fig:InvertedPlaneNoVdWa} shows the results at time $t=20$, and Figure \subref*{fig:InvertedPlaneNoVdWb} at time $t=10^4$. Consistent with the results obtained in \cite{KausBecker} (and references therein), we observe that the viscoelastic film with the highest relaxation time exhibits the fastest dynamics in the development of the RT instability. In fact, in Figure \subref*{fig:InvertedPlaneNoVdWa} the interface of the film with $\la_1 =10$ has significantly developed, while the other films are still in the initial phase of small interfacial amplitude. Eventually, all viscoelastic films reach the near-equilibrium configuration, depicted in Figure \subref*{fig:InvertedPlaneNoVdWb}. A similar behavior was observed by the authors in \cite{BarraEtAl} for the case of regular dewetting processes in the absence of gravity, for which the instability was solely driven by the disjoining attraction/repulsion force with the solid substrate, and for which elasticity was found to facilitate the dynamics of the dewetting process.

\begin{figure}[t]
\captionsetup{type=figure}
\centering
{\includegraphics[width=.95\linewidth,trim=.05cm 0.05cm .0cm 0.05cm,clip=true]{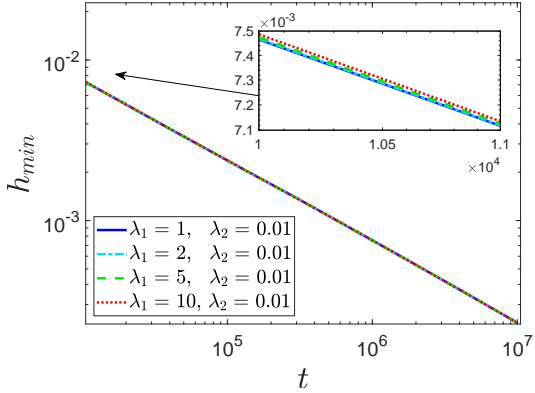}}
\caption{Temporal evolution, in logarithmic scales, of the minimum thickness of viscoelastic films with the same parameters as in (\ref{parameters}) and $\la_2=0.01$, $\la_1 = 1$ (blue solid curve), $\la_1=2$ (cyan dash-dotted curve), $\la_1=5$ (green dashed curve), and $\la_1=10$ (red dotted curve), respectively. The inset shows a magnification for the early times of the evolution, for which $h_{min}$ exhibits an exponential decay. When the disjoining pressure is zero, the thickness reaches zero for times that are long compared to the ones considered in this work.}\label{fig:H_minVsTime}
\end{figure}

\begin{figure}[t]
\captionsetup{type=figure}
\centering
\subfloat[]{\includegraphics[width=.95\linewidth,trim=.0cm 0.05cm .15cm 0.0cm,clip=true]{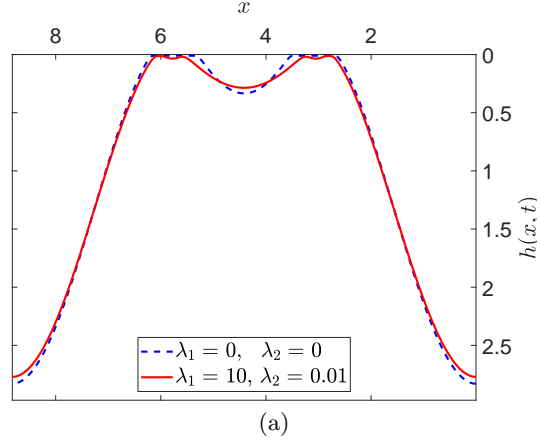}\label{fig:ComparisonViscoelasticityVdWNonZero_a}}\\
\subfloat[]{\includegraphics[width=.95\linewidth,trim=.0cm 0.05cm .15cm 0.0cm,clip=true]{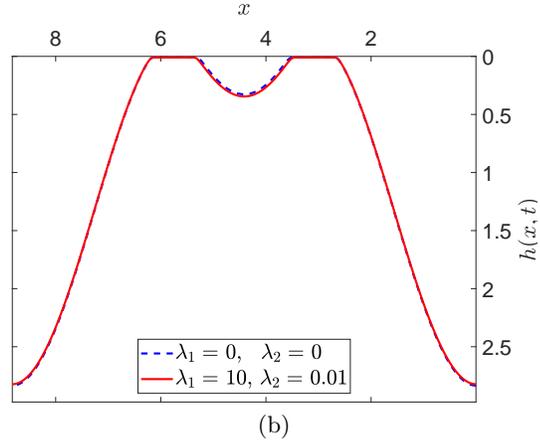}\label{fig:ComparisonViscoelasticityVdWNonZero_b}}\\
\caption{Evolution in the presence of the disjoining pressure, for films with the same parameters as in (\ref{parameters}), and $\la_1 = \la_2 = 0$ (blue dashed curve) compared to a viscoelastic film $\la_1 = 10 ,\,\la_2 = 0.01$ (red solid curve), in \protect\subref{fig:ComparisonViscoelasticityVdWNonZero_a}, at time $t=5.25\times 10^2$, and in \protect\subref{fig:ComparisonViscoelasticityVdWNonZero_b}, at time $t=10^4$.}\label{fig:evolutionVdW}
\end{figure}

Analogously to the results obtained in \cite{BarraEtAl}, we have confirmed here that $\lambda_2$ does not significantly influence the dynamics or the attained morphologies. To investigate this, in Figure \ref{fig:varyingL2}, we plot the evolution of different viscoelastic films with the same parameters as in (\ref{parameters}), in the absence of the disjoining pressure, with $\la_1=1$ and $\la_2 = 0.01$ (blue solid curve with circles), $\la_2=0.1$ (cyan dotted curve with triangles), $\la_2=0.5$ (green dashed curve with squares), and $\la_2=1$ (red solid curve with crosses), respectively. We can see that both in \protect\subref{fig:varyingL2a}, at time $t=1.47\times 10^2$, and in \protect\subref{fig:varyingL2b}, at time $t=10^4$, the profiles of the different film interfaces completely overlap. Therefore, from now on, to investigate the material behavior of the dewetting viscoelastic films, we will focus our attention on varying the relaxation time, rather than the retardation time.

\begin{figure}[t]
\centering
{\includegraphics[width=.9\linewidth,trim=.05cm 0.05cm .05cm 0.05cm,clip=true]{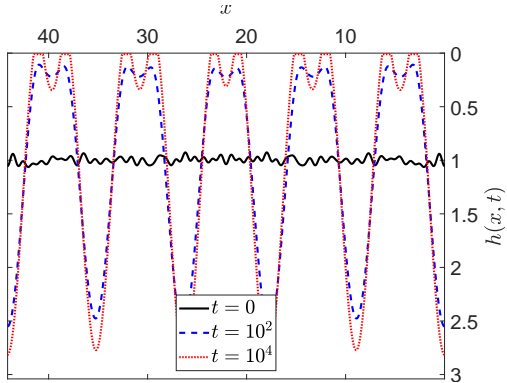}}
\caption{Evolution of a viscoelastic film, with the same parameters as in (\ref{parameters}), and, in particular, $\la_1=5,\la_2=0.01$, initially perturbed by different wavelengths with random amplitude.}\label{fig:5xDom}
\end{figure}

As described earlier, the disjoining pressure induces an equilibrium (precursor) layer on the substrate. Hence, in the absence of this interaction with the substrate, we expect the thickness of the dewetting films to reach zero as $t\rightarrow \infty$. In Figure \ref{fig:H_minVsTime}, we plot the temporal evolution (in logarithmic scales) of the minimum thickness of viscoelastic films in the absence of the disjoining pressure, with the same parameters as in (\ref{parameters}) and $\la_2=0.01$, $\la_1 = 1$ (blue solid curve), $\la_1=2$ (cyan dash-dotted curve), $\la_1=5$ (green dashed curve), and $\la_1=10$ (red dotted curve), respectively. We emphasize that, in Figure \ref{fig:H_minVsTime}, we show the late times of the evolution with $t\in [10^4,10^7]$, for which exponential growth of the interface amplitude is not expected. The slight variations in the behavior of the different viscoelastic films are noticeable only in the (relatively) early times of the evolution shown in the inset of Figure \ref{fig:H_minVsTime}, for which exponential growth of the interface amplitude is observed. Moreover, our results suggest a power law scaling, that is, $h_{min} \propto t^s$, where $s$ is found to be $\sim -1/2$, for the data shown in Figure \ref{fig:H_minVsTime}.

Commonly, only the separate effects of the gravitational force or the disjoining pressure are analyzed. However, for liquid films of micro scale, the crossover region, where the effects of the two forces are comparable, can be considered. Hence, in Figure \ref{fig:evolutionVdW}, we investigate the competing effects of the disjoining pressure and the gravitational force on the dynamics of dewetting films on an inverted plane. We compare the evolutions of two dewetting films: a Newtonian liquid, with $\la_1 = \la_2 = 0$ (blue dashed curve) and a viscoelastic one, with $\la_1 = 10 ,\,\la_2 = 0.01$ (red solid curve), in Figure \subref*{fig:ComparisonViscoelasticityVdWNonZero_a} at time $t=5.25\times 10^2$, and in Figure \subref*{fig:ComparisonViscoelasticityVdWNonZero_b} at time $t=10^4$. Both films form a large droplet in the center of the dewetting region, that is not present when disjoining pressure is not considered (cf.~Figure \subref*{fig:InvertedPlaneNoVdWb} in which only a small hump appears in the hole region). Moreover, in Figure \subref*{fig:ComparisonViscoelasticityVdWNonZero_a}, we notice that the viscoelastic film exhibits two small secondary humps in the interface on the sides of the central droplet. Eventually, in Figure \subref*{fig:ComparisonViscoelasticityVdWNonZero_b}, the small humps disappear, and the viscoelastic films attains a near-equilibrium configuration with a central droplet that is slightly larger than one of the Newtonian film. We emphasize that Figures \ref{fig:InvertedPlaneNoVdW} and \ref{fig:evolutionVdW} include films with the same viscoelastic parameters, i.e., $\la_1=\la_2=0$ (blue curves) and $\la_1=10,\la_2=0.01$ (red curves), and only differ in the absence or presence of the disjoining pressure, respectively. As discussed in \S~\ref{LSAInclinedPlane}, both the disjoining pressure and the gravitational term affect the wavenumbers (and therefore the wavelengths) of maximum instability. However, for the cases presented in Figures \ref{fig:InvertedPlaneNoVdW} and \ref{fig:evolutionVdW}, the wavelength of maximum growth is not significantly different. For the former, $\Lambda_m=8.89$, and, for the latter, $\Lambda_m=8.83$.

To demonstrate that the morphologies observed so far are independent of the particular initial perturbation and domain size chosen (i.e., for the fastest growing wavelength, $\Lambda_m= 2 \pi / k_m$), we analyze a dewetting viscoelastic film, with the same parameters as in (\ref{parameters}), and, in particular, $\la_1=5,\la_2=0.01$, on a computational domain $\Lambda'= 5 \times \Lambda_m$. The film is initially perturbed by different wavelengths, $\Lambda_i = 2 \Lambda_m / i$, with $i=1,2,\ldots,50$, such that
\begin{align}\label{eq:randomIC}
h_0(x,0) = h_0  + \delta h_0 \sum_{i=1}^{50} A_i \cos (2 \pi x / \Lambda_i)  \, ,
\end{align}
\noi where the amplitudes $A_i$ are randomly chosen in the range $[-1,1]$. In Figure \ref{fig:5xDom}, we plot the film evolution, at three different times: $t=0$ (black solid curve), $t=10^2$ (blue dashed curve), and $t=10^4$ (red dotted curve). We can see that, at $t=10^4$, a pattern is formed by satellite droplets that are separated by an average distance $\bar{d}\approx 8.80$, which is close to $\Lambda_m=8.83$, for this set of parameters. Moreover, we notice that, as in the case in which a single wavelength was considered, the droplets appear to reach a steady configuration, during the late times of observation. We remark that, for the physical parameters chosen, the growth rate of the fastest growing wavelength, $\omega_m\approx 0.14$. Therefore, the largest time of evolution plotted in Figure \ref{fig:5xDom} corresponds to $t=10^4 \approx 10^3 \omega_m^{-1}$, that is much longer than the breakup time, comparable to $\omega_m^{-1}$ \cite{MichaelLam2}. A further analysis of the evolution of the satellite droplets is provided in \S~\ref{DropletsAnalysis}.

\begin{figure*}[t]
\centering
\subfloat[]{\includegraphics[width=7cm,valign=t,trim=0cm 0cm 0cm 0.6cm,clip=true]{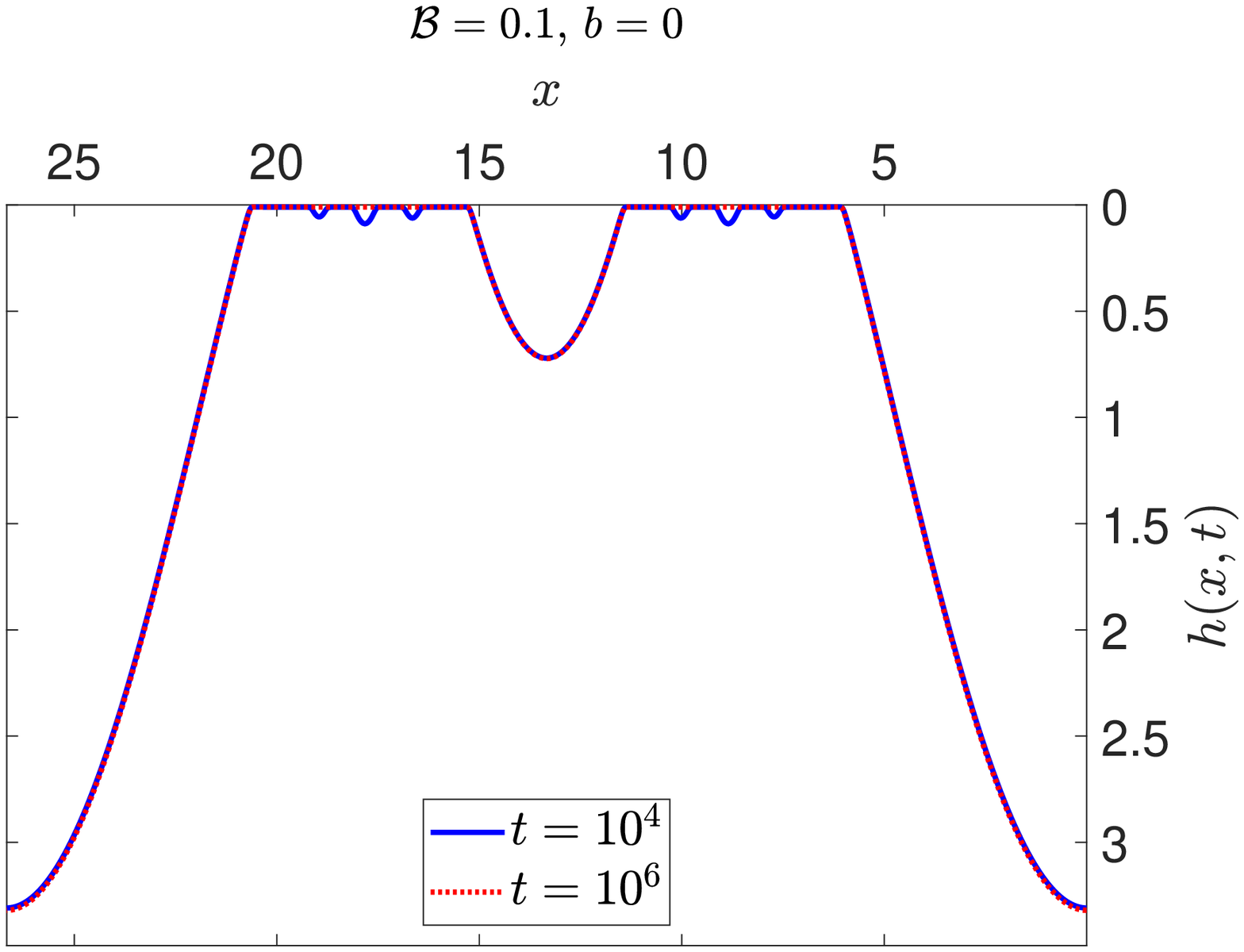}\label{fig:B01,b0_10^4&10^6}}\hspace{1mm}
\subfloat[]{\includegraphics[width=7cm,valign=t,trim=0cm 0cm 0cm 0.6cm,clip=true]{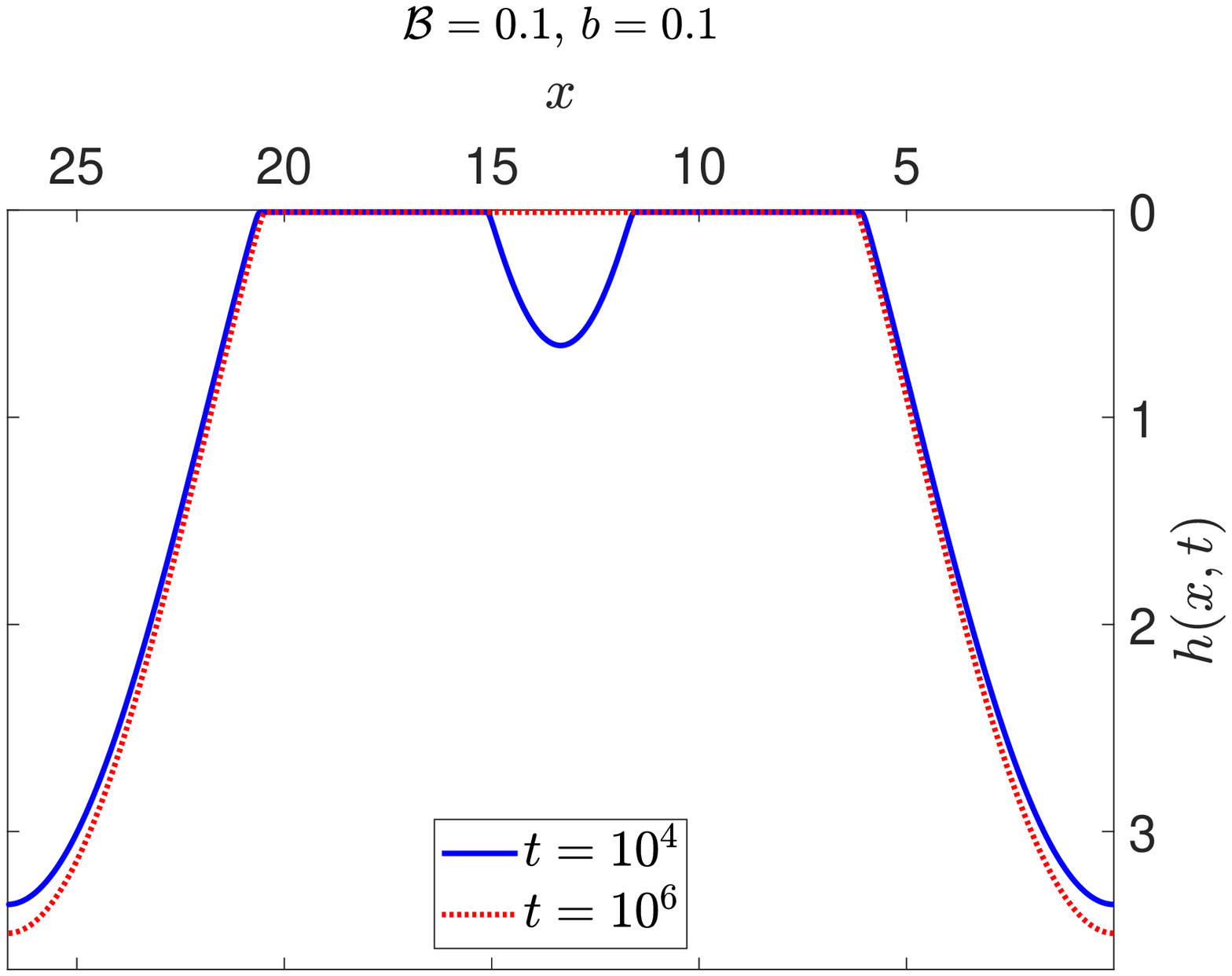}\label{fig:B01,b01_10^4&10^6}}\\
\subfloat[]{\includegraphics[width=7cm,valign=t,trim=0cm 0cm 0cm 0.6cm,clip=true]{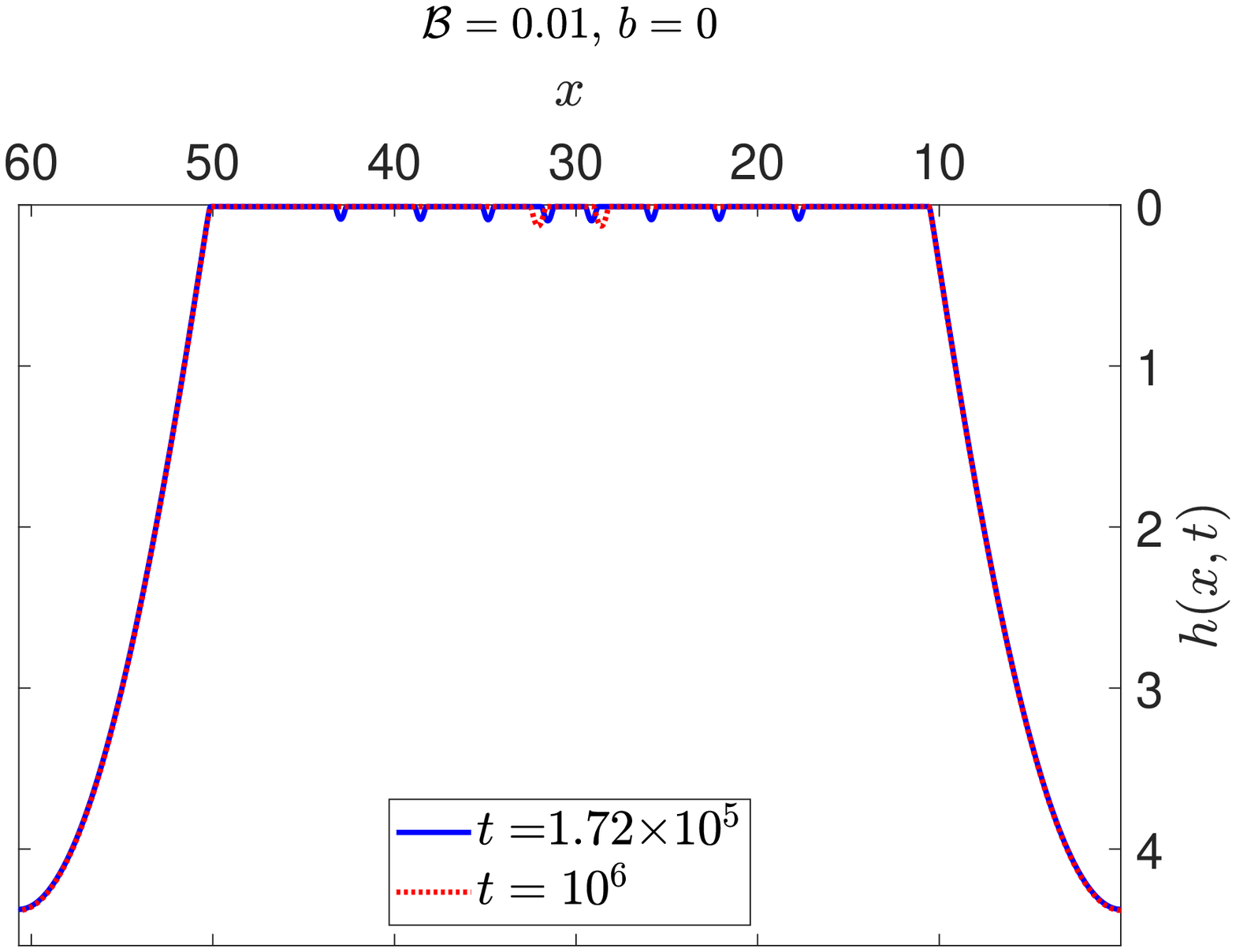}\label{fig:B001,b0_510^5&10^6}}\hspace{1mm}
\subfloat[]{\includegraphics[width=7cm,valign=t,trim=0cm 0cm 0cm 0.6cm,clip=true]{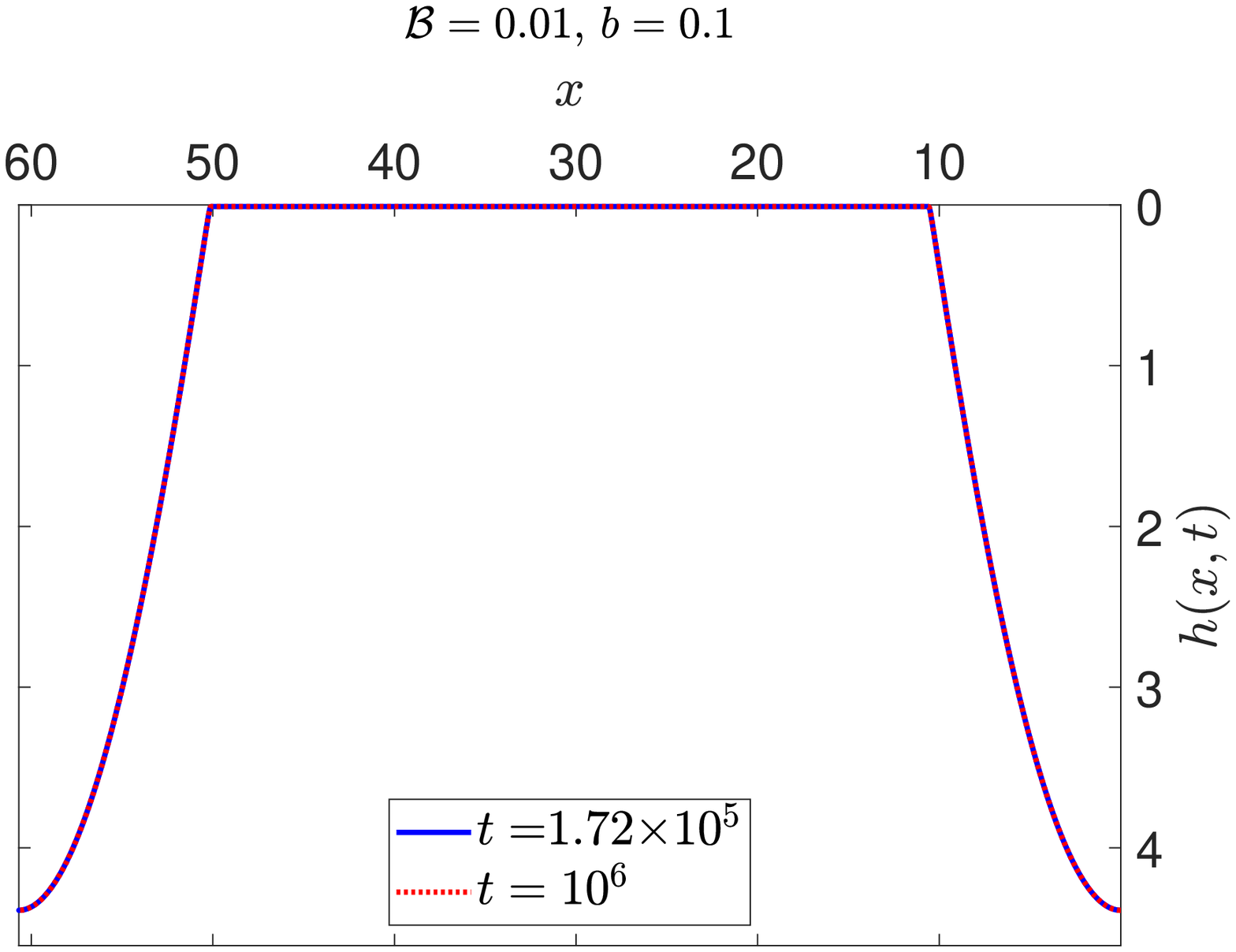}\label{fig:B001,b01_510^5&10^6}}
\caption{Evolution in the presence of the disjoining pressure, for a viscoelastic film with $\la_1 = 10$ and $\la_2 = 0.01$, for $\mathcal{B}=0.1$ (top row), $\mathcal{B}=0.01$ (bottom row), $b=0$ (first column), and $b=0.1$ (second column); in \protect\subref{fig:B01,b0_10^4&10^6} and \protect\subref{fig:B01,b01_10^4&10^6} for $t=10^4$ (blue solid curve) and $t=10^6$ (red dotted curve), and in \protect\subref{fig:B001,b0_510^5&10^6} and \protect\subref{fig:B001,b01_510^5&10^6} for $t=1.72 \times 10^5$ (blue solid curve) and $t=10^6$ (red dotted curve). For these simulations, except for $\mathcal{B}$ and $b$ that have varied, we have kept the fixed parameters presented in (\ref{parameters}).}\label{fig:ComparisonOfBAndb}
\end{figure*}

Next, we investigate the effect of reducing the Bond number, $\mathcal{B}$, on the morphologies of the dewetted films. While $\mathcal{B}$ could be modified in a number of different ways, for the present purposes we may consider that variations of $\mathcal{B}$ are due to change of $g$, e.g., we could consider thin films under microgravity conditions. In Figure \ref{fig:ComparisonOfBAndb}, we plot the evolution of dewetting for a viscoelastic film with $\la_1=10$ and $\la_2 = 0.01$, in the presence of the disjoining pressure, for $\mathcal{B}=0.1$ (top row), and $\mathcal{B}=0.01$ (bottom row). In the results shown so far, we have considered the no-slip boundary condition with the solid substrate (given by $b=0$). We now add to our analysis the effect of slippage, by varying, in Figure \ref{fig:ComparisonOfBAndb}, the slip coefficient from $b=0$ (first column) to $b=0.1$ (second column). In Figure \subref*{fig:B01,b0_10^4&10^6}, for $\mathcal{B}=0.1$ and $b=0$, at time $t= 10^4$, we notice three small secondary droplets in the interface on each side of the main central drop, that eventually, at time $t = 10^6$, disappear. In Figure \subref*{fig:B01,b01_10^4&10^6}, for $\mathcal{B}=0.1$ and $b=0.1$, we can see that, at time $t= 10^4$ only one central drop is present. In fact, the small droplets observed in \subref*{fig:B01,b0_10^4&10^6} are flattened by slippage. However, in the steady configuration attained at time $t=10^6$, the slip with the substrate causes also the central drop to be completely suppressed. We continue our investigation of the microgravity conditions, by further reducing the Bond number. In Figure \subref*{fig:B001,b0_510^5&10^6}, for $\mathcal{B}=0.01$ and $b=0$, at time $t=1.72 \times 10^5$, we can see the formation of multiple secondary droplets, that mostly coalesce in the near-equilibrium configuration, attained at time $t=10^6$. In fact, we can see that only two satellite droplets remain in the final configuration visualized. Finally, in Figure \subref*{fig:B001,b01_510^5&10^6}, we plot the dewetting film for $\mathcal{B}=0.01$, $b=0.1$. By comparing the behavior of this viscoelastic film with the one shown in Figure \subref*{fig:B001,b0_510^5&10^6}, that considers the same times of evolution ($t=1.72 \times 10^5$ and $t=10^6$), we can see how the slippage with the substrate suppresses the satellite droplets. Moreover, we notice how the near-equilibrium configuration displayed in Figure \subref*{fig:B001,b01_510^5&10^6}, at time $t=10^6$ is the same as the one attained at time $t=10^4$. We remark that, for the simulations shown in Figure \ref{fig:ComparisonOfBAndb}, $\mathcal{B}$ and $b$ are the only two parameters varied, and all other parameters in (\ref{parameters}) are fixed.

\subsection[Droplets Analysis]{Droplets Analysis}\label{DropletsAnalysis}

\begin{figure}[t]
\centering
{\includegraphics[width=.9\linewidth,trim=.05cm 0.05cm .05cm 0.05cm,clip=true]{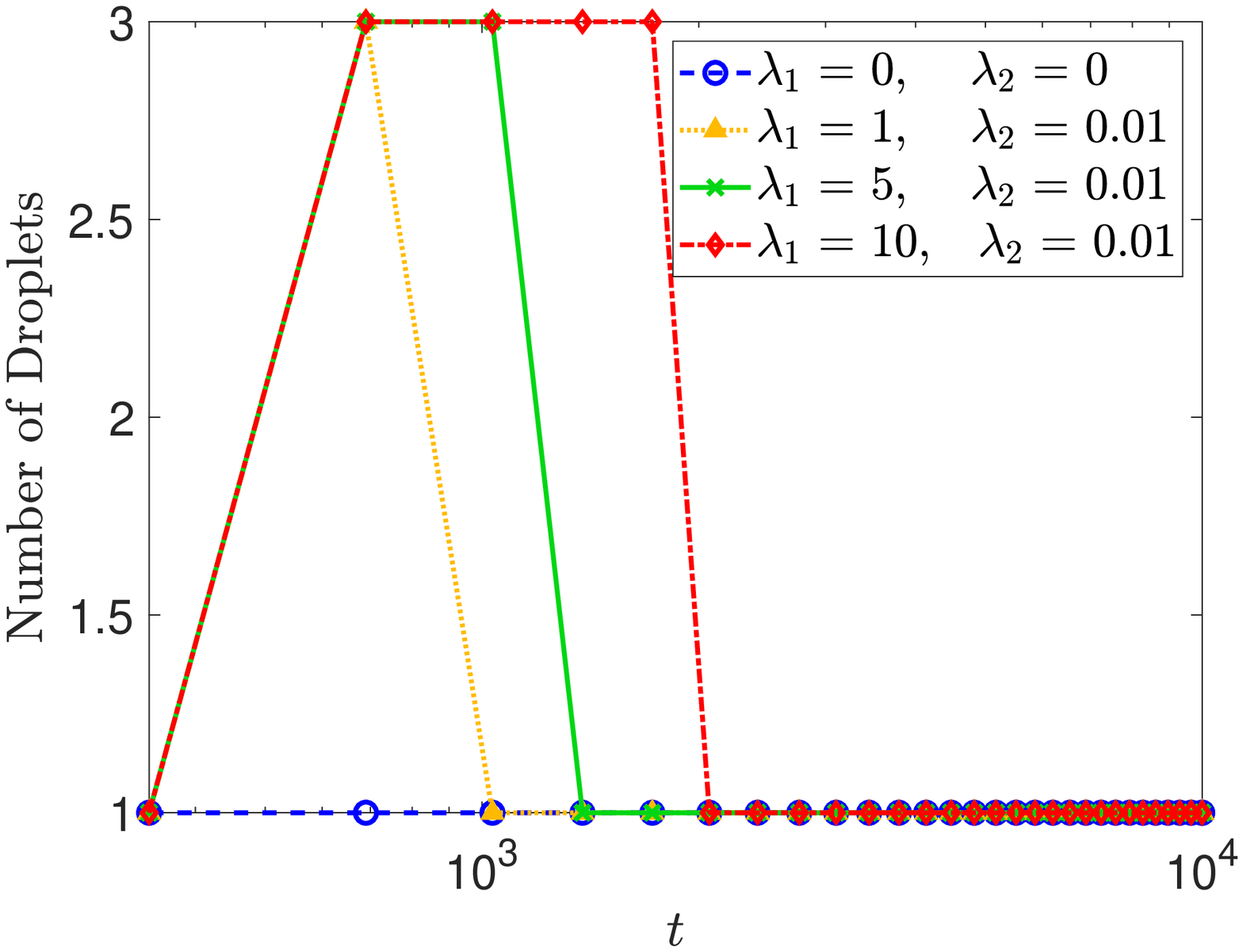}}
\caption{Evolution of the number of droplets, in a semilogarithmic scale, for different films in the presence of the disjoining pressure, with the same parameters as in (\ref{parameters}) and $\la_1 = \la_2 = 0$ (blue circles), $\la_2 = 0.01$ and $\la_1 = 1$ (yellow triangles), $\la_1=5$ (green crosses), and $\la_1=10$ (red diamonds).}\label{fig:DropletsViscoelasticComparison}
\end{figure}

The emergence and evolution of the secondary droplets in thin dewetting films has been of interest in the literature (see, e.g., \cite{BarraEtAl,DiezKondic2007,MichaelLam2}). The secondary length scales of instabilities, that are observed in the nominally dry region between the two separating rims, are induced by the interaction force with the solid substrate (as deduced by comparing the viscoelastic films with same material parameters in Figures \ref{fig:InvertedPlaneNoVdW} and \ref{fig:evolutionVdW}). Moreover, at parity of parameters related to both viscoelasticity and disjoining pressure, these secondary droplets are found to be favored by regimes of microgravity and suppressed by higher slippage with the substrate (as shown in Figure \ref{fig:ComparisonOfBAndb}). We present here an analysis of the dependence of the satellite droplets on the different physical parameters involved. We emphasize that, for the parameter studies that follow, we vary one of the physical quantities at a time and keep all others at their default value, given by (\ref{parameters}).

In Figure \ref{fig:DropletsViscoelasticComparison}, we plot the evolution of the number of satellite droplets, for films with $\la_1 = \la_2 = 0$ (blue circles), $\la_2 = 0.01$ and $\la_1 = 1$ (yellow triangles), $\la_1=5$ (green crosses), and $\la_1=10$ (red diamonds), respectively, in a semilogarithmic scale (linear scale for the $y$-axis and logarithmic scale for the $x$-axis). We can see that viscoelastic films exhibit a higher number of droplets compared to the Newtonian one, and that these secondary instabilities remain longer for higher values of the relaxation time. Eventually these droplets coalesce and reach a steady state (for the time of observation of the current numerical experiments, $t \gg \omega_{m}^{-1}$).

\begin{figure}[t]
\centering
{\includegraphics[width=.9\linewidth,trim=.05cm 0.05cm .05cm 0.05cm,clip=true]{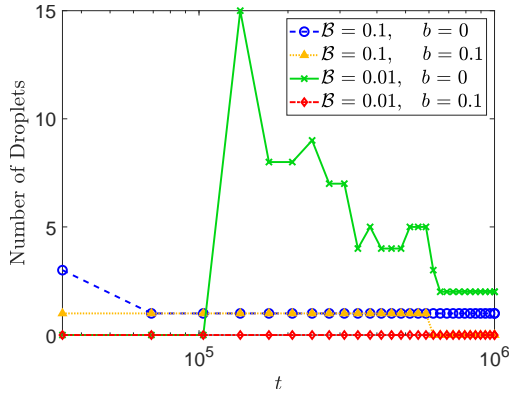}}
\caption{Evolution of the number of droplets, in a semilogarithmic scale, for different films with the same parameters as in Figure \ref{fig:ComparisonOfBAndb}.}\label{fig:DropletsBondAndBComparison}
\end{figure}

In Figure \ref{fig:DropletsBondAndBComparison}, we plot the number of droplets in time, for the same regimes considered in Figure \ref{fig:ComparisonOfBAndb}, in a semilogarithmic scale. We can see how in microgravity conditions, a higher number of droplets is formed, and how, for higher values of slippage with the substrate, these secondary drops are suppressed.

\begin{figure}[t]
\captionsetup{type=figure}
\centering
\subfloat[]{\includegraphics[width=.95\linewidth,trim=.0cm 0.05cm .15cm 0.0cm,clip=true]{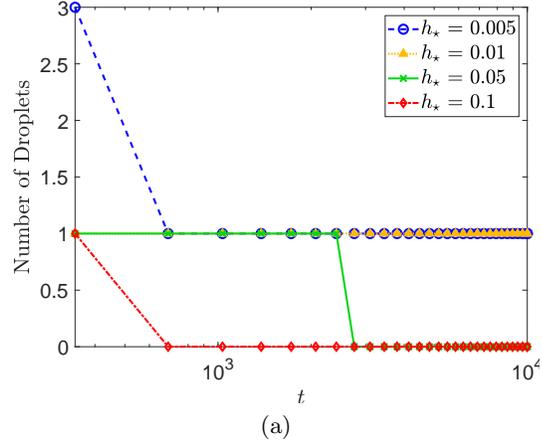}\label{fig:DropletsHstarComparisonNewtonian}}\\
\subfloat[]{\includegraphics[width=.95\linewidth,trim=.0cm 0.05cm .15cm 0.0cm,clip=true]{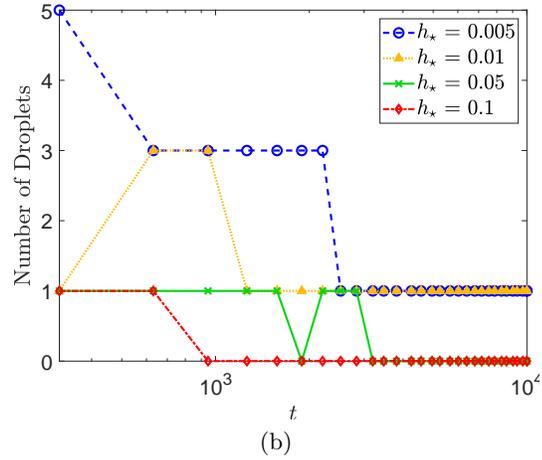}\label{fig:DropletsHstarComparisonViscoelastic}}\\
\caption{Evolution of the number of droplets, in a semilogarithmic scale, for different equilibrium thicknesses: $h_{\star}=0.005$ (blue circles), $0.01$ (yellow triangles), $0.05$ (green crosses), $0.1$ (red diamonds); for a Newtonian film, in \protect\subref{fig:DropletsHstarComparisonNewtonian}, and a viscoelastic one, with $\la_1=5$ and $\la_2=0.01$, in \protect\subref{fig:DropletsHstarComparisonViscoelastic}.}\label{fig:DropletsHstarComparison}
\end{figure}

We proceed by analyzing the effects of the equilibrium (precursor) thickness, $h_{\star}$, induced by the disjoining pressure. In Figure \ref{fig:DropletsHstarComparison}, we plot the evolution of the number of droplets for a Newtonian film with different equilibrium thicknes\-ses: $h_{\star}=0.005$ (blue circles), $0.01$ (yellow triangles), $0.05$ (green crosses), $0.1$ (red diamonds), in a semilogarithmic scale, for a Newtonian film in Figure \subref*{fig:DropletsHstarComparisonNewtonian}, and a viscoelastic one with $\la_1=5$ and $\la_2=0.01$, in Figure \subref*{fig:DropletsHstarComparisonViscoelastic}. We can see how, in both cases, a higher equilibrium thickness suppresses the formation of satellite droplets and favors their coalescence.

\begin{figure}[t]
\captionsetup{type=figure}
\centering
\subfloat[]{\includegraphics[width=.95\linewidth,trim=.0cm 0.05cm .15cm 0.0cm,clip=true]{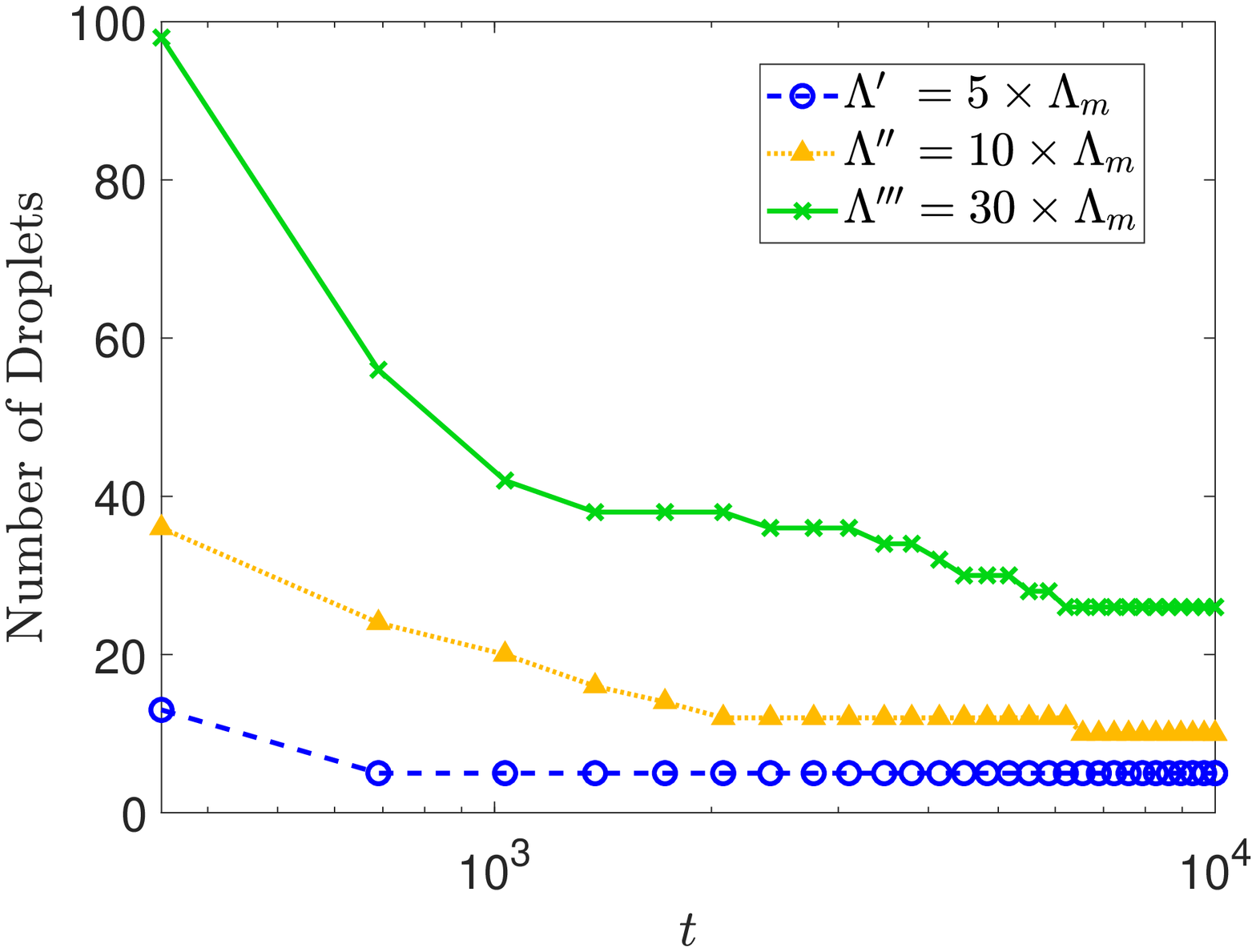}\label{fig:DropletsNumberDifferentDomains}}\\
\subfloat[]{\includegraphics[width=.95\linewidth,trim=.0cm 0.05cm .15cm 0.0cm,clip=true]{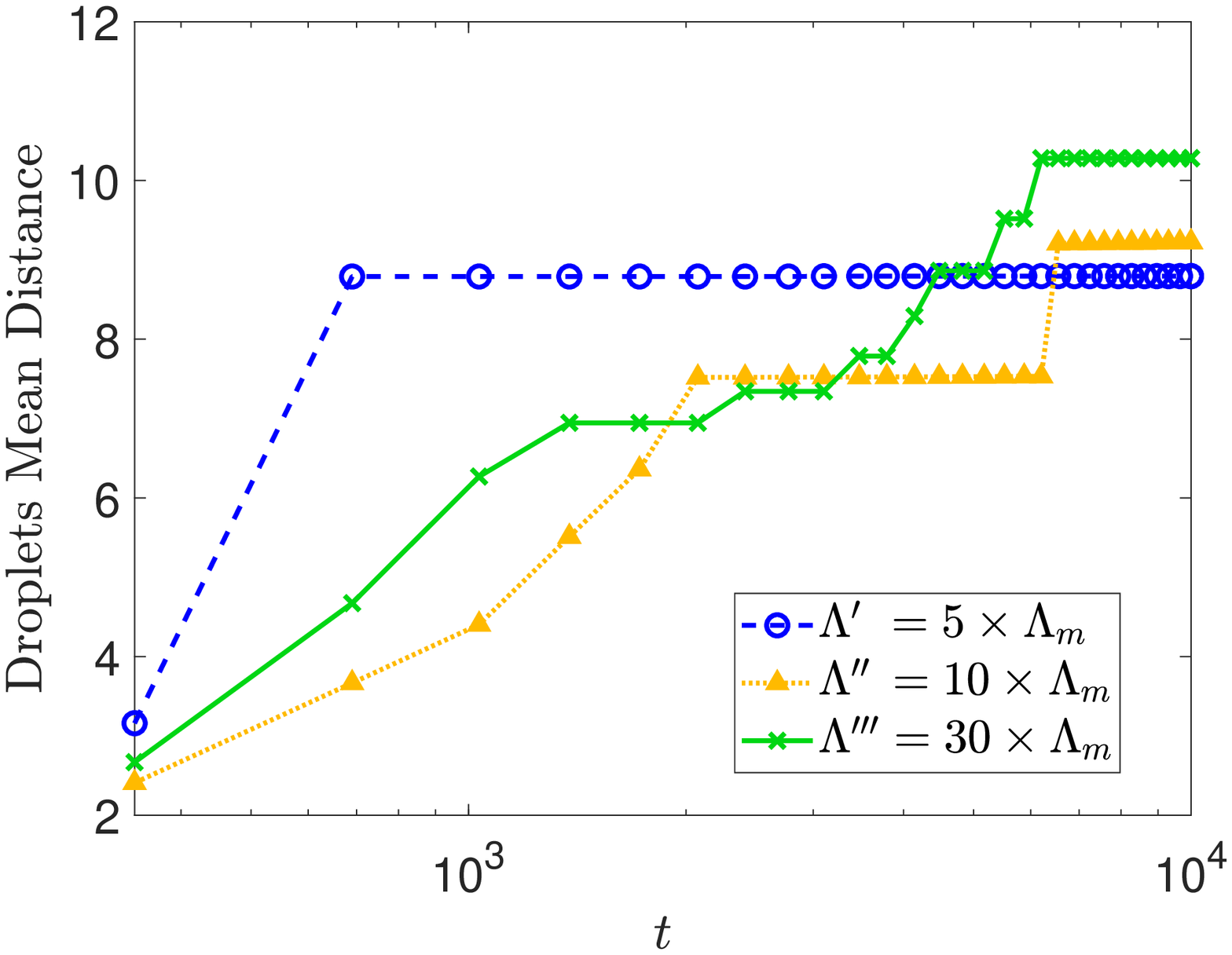}\label{fig:DropletsDistanceDifferentDomains}}\\
\caption{Droplets evolution, in a semilogarithmic scale, for a viscoelastic film with the same parameters as in (\ref{parameters}) and $\la_1=5$ and $\la_2=0.01$, on different computational domains: $\Lambda'=5 \times \Lambda_m$ (blue circles), $\Lambda{''}=10 \times \Lambda_m$ (yellow triangles), and $\Lambda{'''}=30 \times \Lambda_m$ (green crosses); in \protect\subref{fig:DropletsNumberDifferentDomains}, the number of droplets in time; in \protect\subref{fig:DropletsDistanceDifferentDomains}, their mean distance. For these simulations, an initial perturbation with different wavelengths with random amplitude, as in (\ref{eq:randomIC}), is considered.}\label{fig:DifferentDomains}
\end{figure}

The results shown so far, in \S~\ref{DropletsAnalysis}, concerned dewetting films with a computational domain equal to the wavelength of maximum growth, i.e., $\Lambda=\Lambda_m$. Finally, we present a quantitative analysis of the evolution of droplets, for the cases in which the films are initially perturbed by a number of wavelengths with random amplitude, as described in equation (\ref{eq:randomIC}), for different computational domains. In Figure \ref{fig:DifferentDomains}, we consider a dewetting viscoelastic film, with $\la_1=5,\la_2=0.01$, for $\Lambda'=5 \times \Lambda_m$ (blue circles), $\Lambda{''}=10 \times \Lambda_m$ (yellow triangles), and $\Lambda{'''}=30 \times \Lambda_m$ (green crosses). In Figure \subref*{fig:DropletsNumberDifferentDomains}, we plot the evolution of the number of droplets, and, in \subref*{fig:DropletsDistanceDifferentDomains}, their mean distance, both in a semilogarithmic scale. We can see that the maximum number of secondary droplets, $\mathcal{N}_{max}$, varies approximately linearly with the computational domain. In fact, we can compare the results shown in Figure \subref*{fig:DropletsNumberDifferentDomains}, with the ones for the same viscoelastic film for which the computational domain is equal to the single wavelength of maximum growth rate, $\Lambda=\Lambda_m=8.83$ (for this set of parameters), depicted in Figure \ref{fig:DropletsViscoelasticComparison}. We notice that, for $\Lambda=\Lambda_m$, the maximum number of droplets observed is $\mathcal{N}_{max}=3$; for $\Lambda'= 5 \times \Lambda_m'$, $\mathcal{N}_{max}=13$, for $\Lambda{''}= 10 \times \Lambda_m$, $\mathcal{N}_{max}=36$, and, for $\Lambda{'''}= 30 \times \Lambda_m$, $\mathcal{N}_{max}=98$. As discussed in \S~\ref{sec:ThinFilmsResultsInclinedPlane}, we notice that, for all the different computational domains, the number of drop\-lets plateaus to a constant value for large times, i.e., $t\gg \omega_m^{-1}$, indicating that the coalescence of droplets reaches a near-equilibrium state. We emphasize that for times longer than the ones considered in this work, i.e., for $t\rightarrow \infty$, the number of droplets slowly decays to zero (see Glasner and Witelski \cite{GlasnerWitelski}). In our results, the reached minimum number of droplets, $\mathcal{N}_{min}$, also appears to depend linearly on the domain size. In fact, for $\Lambda=\Lambda_m$, $\mathcal{N}_{min}=1$, for $\Lambda'= 5 \times \Lambda_m'$, $\mathcal{N}_{min}=5$, for $\Lambda{''}= 10 \times \Lambda_m$, $\mathcal{N}_{min}=10$, and for $\Lambda{'''}= 30 \times \Lambda_m$, $\mathcal{N}_{min}=26$. Moreover, the mean distance between the droplets is comparable to $\Lambda_m=8.83$ (as displayed in Figure \subref*{fig:DropletsDistanceDifferentDomains}), except for the $\Lambda{'''}$ case, in which the mean distance, $\bar{d}\approx 10.3$.

\section[Conclusions]{Conclusions}\label{sec:InclinedThinFilmsConclusions}

We have considered a novel long-wave governing equation for the interfacial flow of two-dimen\-sional thin viscoelastic films dewetting substrates (with a zero or weak slippage) that can be inclined with respect to the base, under the effects of the gravitational force and the disjoining pressure, in which the stresses are described by the Jeffreys model. We have carried out the linear stability analysis, that shows that the viscoelastic parameters and the slippage coefficient do not influence either the wavenumber corresponding to the maximum growth rate or the critical one. However, the length scales of instabilities are found to be affected by the gravitational contribution. Our numerical results of the computed growth rates in the linear regime are shown to be in agreement with the theoretical prediction given by the linear stability analysis.

We have provided numerical simulations of thin viscoelastic dewetting films in the particular case in which they hang on inverted planes (i.e., for $\alpha = \pi$). To isolate the effects of the gravitational force, we have begun our investigation by analyzing dewetting films in the absence of the disjoining pressure. The results for the nonlinear regime show that, consistent with previous results that did not consider gravity effects \cite{BarraEtAl,TomarEtAl}, a higher relaxation time speeds up the dewetting dynamics.

Furthermore, we have investigated the competing effects of the gravitational and the attraction/repulsion forces, by considering the evolution of viscoelastic films, dewetting an inverted substrate, in the presence of the disjoining pressure. We have found that, at parity of gravitational force (i.e., for the same Bond number), the disjoining pressure induces the formation of satellite droplets. These secondary instabilities, favored by small values of the Bond number, are suppressed when a higher slippage with the substrate is considered. Moreover, we have analyzed the influence of the different physical parameters on the formation and coalescence of the satellite droplets. We have found that a higher value of the equilibrium film thickness suppresses the formation of the secondary instabilities.

In addition, we have verified that our results are independent of the particular initial perturbation and domain size chosen. In fact, by considering dewetting films on a computational domain much longer than the fastest growing wavelength, and by perturbing them with different wavelengths possessing random amplitudes, we have demonstrated that the mean distance between the droplets is accurately described by the wavelength of fastest growth. Finally, we have observed that the number of satellite droplets and their distance scale approximately linearly with the domain size.

Future work shall consider the extension of this numerical investigation in which inclined planes of arbitrary angle $\alpha$ are considered. In that case, boundary conditions for which a constant influx is maintained at the inlet, as in \cite{TeSheng}, would have to be included. Moreover, for arbitrary values of $\alpha$, the form of the perturbation employed for the linear stability analysis should present a wave-like mode \cite{Kondic2003,TeSheng}, rather than an oscillatory one. Furthermore, extensions of this investigation to three spatial dimensions would allow one to describe and capture fingering instabilities, known to arise in the direction transversal to the flow \cite{Kondic2003,DeBruyn}.

\section{Acknowledgments}
The authors would like to thank the referee for the very thorough and detailed revision provided.

\section{Authors contributions}
All the authors were involved in the preparation of the manuscript.
All the authors have read and approved the final manuscript.

\numberwithin{equation}{section}
\section*{Appendix: Derivation}
\renewcommand{\theequation}{A.\arabic{equation}}
\setcounter{equation}{0}  

We define the outward unit normal, $\mathbf{n}$, as
\begin{align}\label{nt}
\mathbf{n}&=\frac{1}{{\left(\left(h_x\right)^2 +1\right)}^{1/2}}\left(-h_x, 1 \right) \, .
\end{align}
\noindent The kinematic boundary condition is given by $ Df/Dt = f_t + \textbf{v} \cdot \nabla f = 0$ (where we have used the material derivative $D(\cdot)/Dt$), and in which substituting $f(x,y,t)=y-h(x,t)$ gives
\begin{equation}\label{ht}
h_t (x,t) = - \frac{\partial}{\partial x} \int_{0}^{h(x,t)} v_1(x,y) dy \, .
\end{equation}
\noindent As anticipated, the boundary conditions at the solid substrate are described by the non-penetra\-tion condition for the normal component of the velocity and the Navier slip boundary condition for the tangential one, respectively
\begin{equation}\label{Eq:BottomBC}
v_2=0, \quad v_1= \frac{b}{\eta}\sigma_{12} \, ,
\end{equation}
\noindent where for the Navier slip condition we use the Newtonian shear stress, that is dominant in regions away from the contact line \cite{QuianEtAl}. We observe that $b=0$ implies a no-slip boundary condition with the substrate. The influence of this parameter on the morphology of the dewetting front in lubrication models has been investigated both theoretically and experimentally, for instance see \cite{Fetzer,Munch}.

We nondimensionalize the system of governing equations using common scalings for long-wave formulations
\begin{align}
&x=Lx^* \, , \; (y,h,h_{\star},b)=H(y^*,h^*,h_{\star}^*,b^*)\, ,  \label{NonDim1}\\
&v_1=Vv_1^* \, , \; v_2=\varepsilon V v_2^* \, , \; (p,\Pi)= P(p^*,\Pi^*) \, , \label{NonDim2}\\
&(t,\la_1,\la_2)=T( t^*,\la_1^*,\la_2^*) \, , \; \label{NonDim3}
\end{align}
\begin{equation}\label{NonDimTensor}
\left(
\begin{array}{cc}
\sigma_{11}&\sigma_{12}\\
\sigma_{12}&\sigma_{22}\\
\end{array}
\right)
=\frac{\eta}{T}
\left(
\begin{array}{cc}
\sigma_{11}^*&\frac{\sigma_{12}^*}{\vare}\\
\frac{\sigma_{12}^*}{\vare}&\sigma_{22}^*\\
\end{array}
\right) \, ,
\end{equation}
\noindent where ${H}/{L}= \varepsilon \ll 1$ is the small parameter. To balance pressure, viscous and capillary forces, the pressure is scaled with $P= {\eta}/({T \varepsilon^2})$, the surface tension with $\gamma = \Gamma \gamma^*$, where $\Gamma = {V \eta}/{\vare^3} $, and the time with $T = L/ V$. Following the formulation in \cite{Rauscher2005}, we can consider the scaled surface tension, $\gamma^*$, to be equal to one, by choosing the velocity so that the inverse capillary number is $\textrm{Ca}^{-1}=\vare^3\gamma / (V\eta)=1$ \cite{Myers}, and, subsequently, $L = \gamma \vare^3 T / \eta$. Moreover, we notice that, by choosing the vertical length scale, $H$, to be equal to the initially flat fluid interface height, we can simplify the dispersion relations in \S~\ref{LSAInclinedPlane}, with the value $h_0 =1$. However, we refrain from applying this simplification to allow for generalization to arbitrary choices of the reference film thickness.

We note that one could also consider other scaling factors. For instance, if gravity is considered to be the sole driving force for instabilities, one could set the Bond number $\mathcal{B}=1$. This leads to the capillary length scale, i.e., $L = (\gamma / (\rho g))^{1/2}$. Subsequently, the pressure would be scaled by $P= \gamma H / L^2 = \rho g H$, the time by $T = \eta \gamma / H^3 (\rho g)^2$, the velocity determined by $V=L/T$, and the normalized surface tension, similarly to the former case, would be $\gamma^{*}=1$. However, in the current work, we are interested in the competing mechanisms of the disjoining pressure and the gravitational force that together drive the instabilities and affect their length scales; moreover, as discussed in \S~\ref{sec:ThinFilmsResultsInclinedPlane}, we want to be able to vary the Bond number to be able to analyze the behavior of films under the microgravity conditions. For this reason, we prefer the former choice of scalings, given by equations (\ref{NonDim1})-(\ref{NonDimTensor}). To avoid a cumbersome notation, we drop the superscript `$^*$' and consider for the rest of this document all quantities to be dimensionless.

The incompressibility condition (\ref{IncompressibilityChapter3}) is invariant under rescalings, while the dimensionless forms of equation (\ref{GovMomentumChapter3}) for the $x$ and $y$ components, respectively, are
\begin{subequations}\label{Eq:x-y-compMomentumWithGravity}
\begin{align}
\varepsilon^2 Re \frac{D v_1}{Dt}&= \varepsilon^2 \frac{\partial \sigma_{11}}{\partial x} +\frac{\partial \sigma_{12}}{\partial y} - p_x + \mathcal{S} \, , \label{Eq:x-compMomentumWithGravity}\\
\varepsilon^4 Re \frac{D v_2}{Dt}&= \varepsilon^2 \left( \frac{\partial \sigma_{12}}{\partial x}  + \frac{\partial \sigma_{22}}{\partial y}
 \right) - p_y - \mathcal{C} \, , \label{Eq:y-compMomentumWithGravity}
\end{align}
\end{subequations}
\noindent where $Re= {\rho V L}/{\eta}$ is the Reynolds number, assumed to be of order $1 / \varepsilon$ or smaller. In equations (\ref{Eq:x-y-compMomentumWithGravity}) we have used $ \mathcal{S}$ and $ \mathcal{C}$, whose definitions are given in equations (\ref{Def:S&C}), and the extended notation for the derivatives \mbox{$\partial (\cdot) / \partial x$} and \mbox{$\partial (\cdot) / \partial y$} (this version will be used from now on, whenever needed to avoid double subscripts). The dimensionless distinct components of the stress tensor given by the Jeffreys model, equation (\ref{Jeffreys}), satisfy
\begin{subequations}\label{nondimJeffrey}
\begin{align}
\left( 1 + \la_1 \partial_t \right)\sigma_{11}  &= 2 \left( 1 + \la_2 \partial_t \right) \frac{\partial v_1}{\partial x}  \, , \\
\left( 1 + \la_1 \partial_t \right)\sigma_{22} &= 2 \left( 1 + \la_2 \partial_t \right) \frac{\partial v_2}{\partial y} \, , \\
\left( 1 + \la_1 \partial_t \right)\sigma_{12} &=  \left( 1 + \la_2 \partial_t \right)\frac{\partial v_1}{ \partial y} + \nonumber \\
\quad & \varepsilon^2 \left( 1 + \la_2 \partial_t \right) \frac{\partial v_2}{ \partial x} \, . \label{nondimJeffrey3}
\end{align}
\end{subequations}
\noindent The kinematic boundary condition, equation (\ref{ht}), is invariant under rescaling, while the non-penetration condition and the Navier slip boundary condition for the velocity components parallel to the substrate, given in equation (\ref{Eq:BottomBC}), in dimensionless form are
\begin{equation}\label{Eq:DimlessBottomBCs}
v_2=0 \, ,\quad v_1= b \sigma_{12} \, ,
\end{equation}
\noindent where in the weak slip regime $b=O(1)$ \cite{Blossey2006}. The leading order terms in the governing equations (\ref{Eq:x-compMomentumWithGravity}) and (\ref{Eq:y-compMomentumWithGravity}), respectively, are
\begin{subequations}\label{Eq:Rescaled-x-y-compMomentumWithGravity}
\begin{align}
\frac{\partial \sigma_{12}}{\partial y} &= p_x- \mathcal{S} \, , \label{Eq:Rescaled-x-compMomentumWithGravity} \\
p_y &= - \mathcal{C} \, . \label{Eq:Rescaled-y-compMomentumWithGravity}
\end{align}
\end{subequations}
\noindent  The leading order terms of the normal and tangential components of the stress balance at the free surface, $y=h(x,t)$, equation (\ref{Eq:LaplacePressBCChapter3}), respectively are
\begin{align}\label{Def:pRWithGravityAtInterface}
p = - h_{xx}  - \Pi\, \textrm{, on } y=h(x,t) \, ,
\end{align}
\noindent and
\begin{align}\label{Def:36}
\sigma_{12} &= 0 \,\textrm{, on } y=h(x,t)\, .
\end{align}
\noindent From equation (\ref{Eq:Rescaled-y-compMomentumWithGravity}) we know that the pressure is a linear function of $y$ (see also \cite{Hsieh2,Myers}). Hence, by integrating equation (\ref{Eq:Rescaled-y-compMomentumWithGravity}) and by using the boundary condition for the pressure at the interface, given by the leading order term of the normal component of the pressure balance, equation (\ref{Def:pRWithGravityAtInterface}), we obtain
\begin{align}\label{Def:pressureWithGravity}
p=  - h_{xx}  - \Pi - \mathcal{C} (y- h) \, .
\end{align}
\noindent The $x$ component of the pressure gradient is given by differentiating equation (\ref{Def:pressureWithGravity}) in the $x$ direction, obtaining
\begin{align}\label{Def:PressureGradientWithGravity}
p_x &= - h_{xxx}  - \Pi'h_x  + \mathcal{C} h_x \, .
\end{align}
\noindent The nondimensional form of $\Pi$ in equations (\ref{Def:pRWithGravityAtInterface}) and (\ref{Def:pressureWithGravity}) is given by
\begin{equation}\label{Def:VdW}
\Pi= \frac{\gamma(1 - \cos \theta_e)}{\vare^2 M h_{\star}} \left[ {\left( \frac{h_{\star}}{h}  \right)}^{m_1} - {\left( \frac{h_{\star}}{h}\right)}^{m_2} \right] \, ,
\end{equation}
\noindent where, we remark that all the quantities are considered to be normalized. Moreover, in order for the expression in equation (\ref{Def:VdW}) to be $O(1)$, we expand $\cos \theta_e$, by considering $\theta_e = \vare \theta_e^*$, so that $1 - \cos\theta_e \approx \vare^2 {\theta_e^*}^2/2$. Thus, we can recast
\begin{equation}
\label{LinearizedVdW}
\Pi \approx \frac{\gamma {\theta_e^*}^2}{2 M h_{\star}} \left[ {\left( \frac{h_{\star}}{h}  \right)}^{m_1} - {\left( \frac{h_{\star}}{h}\right)}^{m_2} \right] \, .
\end{equation}
\noindent For the particular set of parameters considered in the present study (see \S~\ref{sec:MathematicalFormulationInclinedPlane} and (\ref{parameters})), we notice that $\gamma {\theta_e^*}^2 / 2 M h_{\star} \sim \gamma(1 - \cos \theta_e) / M h_{\star} = O(1)$ (where the latter expression has been used in this work).

Integrating equation (\ref{Eq:Rescaled-x-compMomentumWithGravity}) from $y$ to $h(x,t)$, we obtain
\begin{align}\label{Eq:tau_21WithGravity}
\sigma_{12}= (y -h) p_x  -(y-h)\mathcal{S}\, .
\end{align}
\noindent Substituting this form of $\sigma_{12}$, equation (\ref{Eq:tau_21WithGravity}), into equation (\ref{nondimJeffrey3}), we obtain (up to the leading order)
\begin{align}\label{Eq:40WithGravity}
&\left( 1 + \la_1 \partial_t \right) \left[ p_x (y-h)  - \mathcal{S} (y-h)  \right] = \nonumber \\
&\left( 1 + \la_2 \partial_t \right)\frac{\partial v_1}{\partial y}  \, .
\end{align}
\noindent Integrating equation (\ref{Eq:40WithGravity}) from $0$ to $y$ and using the corresponding boundary conditions at the substrate, given in equation (\ref{Eq:DimlessBottomBCs}), we obtain
\begin{align}\label{Eq:41WithGravity}
&\left( 1 + \lambda_2 {\partial_t} \right) \left( v_1 + b h p_x -bh\mathcal{S} \right) = \nonumber\\
&\left( 1 + \lambda_1 {\partial_t} \right) \left[ \left( \frac{y^2}{2} - y h \right) \left( p_x - \mathcal{S} \right)\right] \, .
\end{align}
\noindent Integrating equation (\ref{Eq:41WithGravity}) from $y=0$ to $y=h(x,t)$ gives
\begin{align}\label{Eq:42WithGravity}
&\left( 1 + \lambda_2 {\partial_t} \right) \left[\int_0^{h(x,t)} v_1 \,dy + bh^2 p_x -
bh^2\mathcal{S} \right] - \nonumber \\
&\lambda_2 h_t \left( v_1(y=h(x,t)) + bh p_x - bh \mathcal{S}  \right) = \nonumber \\
& \; - \left( 1 + \la_1 \partial_t \right) \left[  \frac{h^3}{3}  \left( p_x  - \mathcal{S} \right) \right]+ \lambda_1 \frac{ h^2}{2} h_t \left( p_x - \mathcal{S}\right) \, .
\end{align}
\noindent Taking the spatial derivative of the latter equation and substituting it into the kinematic boundary condition (\ref{ht}), we obtain a long-wave approximation in terms of $v_1$ and $h(x,t)$
\begin{align}\label{Eq:43WithGravity}
& h_t + \lambda_2 \left[ h_{tt} +  \partial_x({{v_1}} (y=h(x,t)) h_t ) \right] = \nonumber\\
& \partial_x \left[ (1 + \lambda_1 \partial_t) \left( \frac{h^3}{3} p_x - \frac{h^3}{3} \mathcal{S}\right) + \right. \nonumber \\
& \left. \ (1 + \lambda_2 \partial_t ) \left(bh^2 p_x -bh^2\mathcal{S}\right) \right] - \nonumber\\
& \partial_x \left\{ \left[ \lambda_1 \frac{h^2}{2}\left( p_x -\mathcal{S} \right) + \lambda_2 b h \left( p_x -\mathcal{S}\right)\right] h_t \right\} \, .
\end{align}
\noindent To write this in a closed form relation for $h(x,t)$, we note that equation (\ref{Eq:41WithGravity}) can be written in a more compact form as a linear ordinary differential equation for $v_1$ (assuming all other quantities known at a given time), as
\begin{align}\label{Eq:44WithGravity}
&v_1 + \la_2 \frac{\partial v_1}{\partial t} = - (1 + \la_2 \partial_t) \left(bh p_x -bh\mathcal{S} \right) + \nonumber \\
&(1 + \la_1 \partial_t) \left[ \left( \frac{y^2}{2} - hy \right) \left(p_x -\mathcal{S} \right) \right] \, .
\end{align}
\noindent One can simply solve equation (\ref{Eq:44WithGravity}) by integrating in time, obtaining
\begin{equation}\label{Eq:46WithGravity}
v_1 = \frac{1}{\la_2} \int_{- \infty}^{t} e^{- \frac{t - {t'}}{\la_2}} \tilde{f}(x,y,{t'}) d {t'}\, ,
\end{equation}
\noindent with $\tilde{f}$ equal to the right-hand side of equation (\ref{Eq:44WithGravity}). Integration by parts can be performed to recast equation (\ref{Eq:46WithGravity}) at $y=h(x,t)$, and finally one finds the dimensionless form of the governing equation (\ref{Eq:GovEqnWithGravity1D}).

\bibliographystyle{spphys}

\begin{thebibliography}{100}


\bibitem{DeGennes2}
P.-G. de~Gennes,
\newblock {\em Rev. Mod. Phys.} \textbf{57}, 827--863 (1985).

\bibitem{ScardovelliZaleski}
R.~Scardovelli and S.~Zaleski,
\newblock {\em Annu. Rev. Fluid Mech.} \textbf{31}, 567--603 (1999).

\bibitem{TryggvasonEtAl}
G. Tryggvason, R. Scardovelli, and S. Zaleski,
\newblock {\em Direct numerical simulations of gas-liquid multiphase flows}
\newblock (Cambridge University Press, Cambridge, 2011).

\bibitem{Jeffreys}
H. Jeffreys,
\newblock {\em The Earth: its origin, history, and physical constitution}
\newblock (Cambridge University Press, Cambridge, 1952).

\bibitem{Isreaelachvili}
J. Israelachvili,
\newblock {\em Intermolecular \& surface forces}
\newblock (Academic Press, London, 1985).

\bibitem{Bird}
R.A. Bird, R.C. Armstrong, and O. Hassager,
\newblock {\em Dynamics of polymeric liquids: volume $1$ fluid mechanics}
\newblock (Wiley-Interscience, Toronto, 1987).

\bibitem{Reynolds}
O. Reynolds,
\newblock {\em Philos. Trans. R. Soc. London} \textbf{177}, 157--234 (1886).

\bibitem{Oron}
A. Oron, S. Davis, and G. Bankoff,
\newblock {\em Rev. Mod. Phys.} \textbf{69}, 931--980 (1997).

\bibitem{B-W}
F. Brochard-Wyart, G. Debregeas, R. Fondecave, and P. Martin,
\newblock {\em Macromolecules} \textbf{30}, 1211--1213 (1997).

\bibitem{Reiter}
G. Reiter,
\newblock {\em Phys. Rev. Lett.} \textbf{68}, 75--80 (1992).

\bibitem{Safran}
S.A. Safran and J. Klein,
\newblock {\em J. Phys. II} \textbf{3}, 749--757 (1993).

\bibitem{GabrieleEtAl2}
S. Gabriele, S. Sclavons, G. Reiter, and P. Damman,
\newblock {\em Phys. Rev. Lett.} \textbf{96}, 156105 (2006).

\bibitem{Rauscher2005}
M. Rauscher, A. M{\"u}nch, B. Wagner, and R. Blossey,
\newblock {\em Eur. Phys. J. E} \textbf{17}, 373--379 (2005).

\bibitem{Blossey2006}
R. Blossey, A. M{\"u}nch, M. Rauscher, and B. Wagner,
\newblock {\em Eur. Phys. J. E} \textbf{20}, 267--271 (2006).

\bibitem{BlosseyBook}
R. Blossey,
\newblock {\em Thin Liquid Films. Dewetting and Polymer Flow}
\newblock (Springer, New York, 2012).


\bibitem{BarraEtAl}
V. Barra, S. Afkhami, and L. Kondic,
\newblock {\em J. {N}on-{N}ewton. Fluid Mech.} \textbf{237}, 26--38 (2016).

\bibitem{TomarEtAl}
G. Tomar, V. Shankar, S.K. Shukla, A. Sharma, and G. Biswas,
\newblock {\em Eur. Phys. J. E} \textbf{20}, 185--199 (2006).

\bibitem{BenzaquenEtAl}
M. Benzaquen, T. Salez, and E. Rapha\"{e}l,
\newblock {\em EPL (Eur. Lett.)} \textbf{106}, 36003 (2014).

\bibitem{Maxwell}
J.C. Maxwell,
\newblock {\em Phil. Trans. R. Soc. Lond.} \textbf{157}, 49--88 (1867).

\bibitem{Huppert}
H.E. Huppert,
\newblock{\em Nature} \textbf{300}, 427--429 (1982).

\bibitem{Hsieh}
D.-Y. Hsieh,
\newblock {\em Phys. Fluids} \textbf{8}, 1785--1791 (1965).

\bibitem{Schwartz}
L.W. Schwartz,
\newblock {\em Phys. Fluids A: Fluid} \textbf{1}, 443--44 (1989).

\bibitem{KellyGoussis}
R.E. Kelly, D.A. Goussis, S.P. Lin, and F.K. Hsu,
\newblock {\em Phys. Fluids A: Fluid} \textbf{1}, 819--828 (1989).

\bibitem{Hsieh2}
D.-Y. Hsieh,
\newblock {\em Phys. Fluids A: Fluid} \textbf{2}, 1145--1148 (1990).

\bibitem{Kondic2003}
L. Kondic,
\newblock {\em SIAM Rev.} \textbf{45}, 95--115 (2003).

\bibitem{GombaKondic}
J.M. Gomba, J. Diez, R. Gratton, A.G. González, and L. Kondic,
\newblock{\em Phys. Rev. E} \textbf{76}, 046308 (2007).

\bibitem{Kull}
H.J. Kull,
\newblock{\em Phys. Rep.} \textbf{206}, 197--325 (1991).

\bibitem{Kapitza}
S.P. Kapitza,
\newblock{\em Zh. Eksp. Teor. Fiz.} \textbf{18}, 3--28 (1948).

\bibitem{KofmanEtAl}
N. Kofman, W. Rohlfs, F. Gallaire, B. Scheid, and C. Ruyer-Quil,
\newblock{\em Int. J. Multiphas. Flow} textbf{104}, 286--293 (2018).

\bibitem{MichaelLam}
M.A. Lam, L.J. Cummings, T.-S. Lin, and L. Kondic,
\newblock {\em J. Eng. Math.} \textbf{94}, 97--113 (2015).

\bibitem{Tshehla}
M.S. Tshehla,
\newblock {\em Math. Probl. Eng.} \textbf{2013}, 1--8 (2013).

\bibitem{PicchiEtAl}
D. Picchi, P. Poesio, A. Ullmann, and N. Brauner,
\newblock {\em Int. J. Multiphas. Flow} \textbf{97}, 109--133 (2017).

\bibitem{KausBecker}
B.J. Kaus and Becker T.W.,
\newblock{\em Geophys. J. Int.} \textbf{168}, 843--862 (2006).

\bibitem{BarraEtAl2}
V. Barra, S.A. Chester, and S. Afkhami,
\newblock {\em Comp. \& Fluids} \textbf{175}, 36--47 (2018).

\bibitem{MunchWagner}
A. M\"unch, B. Wagner, M. Rauscher, and R. Blossey,
\newblock {\em Eur. Phys. J. E} \textbf{20}, 365--368 (2006).

\bibitem{Larson}
R.G. Larson,
\newblock {\em The structure and rheology of complex fluids}
\newblock (Oxford University Press, Oxford, 1999).

\bibitem{Siginer}
D.A. Siginer,
\newblock {\em Stability of non-linear constitutive formulations for
  viscoelastic fluids}
\newblock (Springer, New York, 2014).


\bibitem{MainardiSpada}
F. Mainardi and G. Spada,
\newblock {\em Eur. Phys. J. Spec. Topics} \textbf{193}, 133--160 (2011).

\bibitem{Gutierrez-Lemini}
D. Gutierrez-Lemini,
\newblock {\em Engineering viscoelasticity}
\newblock (Springer, New York, 2014).

\bibitem{Fetzer}
R. Fetzer, K. Jacobs, A. M{\"u}nch, B. Wagner, and T.P. Witelski,
\newblock {\em Phys. Rev. Lett.} \textbf{95}, 127801 (2005).

\bibitem{Munch}
A. M{\"u}nch, B. Wagner, and T.P. Witelski,
\newblock {\em J. Eng. Math.} \textbf{53}, 359--383 (2005).

\bibitem{DiezKondic2007}
J.A. Diez and L. Kondic,
\newblock {\em Phys. Fluids} \textbf{19}, 072107 (2007).

\bibitem{Teletzke}
G. Teletzke, H.T. Davis, and L.E. Scriven,
\newblock {\em Chem. Eng. Commun.} \textbf{55}, 41--82 (1987).

\bibitem{Ivana}
I. Seric, S. Afkhami, and L. Kondic,
\newblock {\em J. Fluid Mech.} \textbf{755}, 1--12 (2014).

\bibitem{Kyle}
K. Mahady, S. Afkhami, and L. Kondic,
\newblock {\em J. Comput. Phys.} \textbf{294}, 243--257 (2015).

\bibitem{TeSheng}
T.-S. Lin and L. Kondic,
\newblock {\em Phys. Fluids} \textbf{22}, 052105 (2010).

\bibitem{DiezKondic2002}
J.A. Diez and L. Kondic,
\newblock {\em J. Comput. Phys.} \textbf{183}, 274--306 (2002).

\bibitem{Bertozzi}
A.L. Bertozzi,
\newblock {\em Not. Amer. Math. Soc.} \textbf{45}, 689–-697 (1998).

\bibitem{MichaelLam2}
M.A. Lam, L.J. Cummings, T.-S. Lin, and L. Kondic,
\newblock {\em J. Fluid Mech.} \textbf{841}, 925--961 (2018).

\bibitem{GlasnerWitelski}
K.B. Glasner, and  T.P. Witelski,
\newblock {\em Phys. Rev. E} \textbf{67}, 016302 (2003).

\bibitem{DeBruyn}
J.R. De~Bruyn,
\newblock {\em Phys. Review A} \textrm{46}, R4500--4503 (1992).

\bibitem{QuianEtAl}
T. Qian, X.P. Wang, and P. Sheng
\newblock {\em Comm. Math. Sci.} \textbf{1}, 333--341, (2003).

\bibitem{Myers}
T.G. Myers,
\newblock {\em SIAM Rev.} \textbf{3}, 441--462 (1998).

\end{thebibliography}


\end{document}